\documentclass[journal,letterpaper,twoside]{IEEEtran}
\usepackage{cite}
\usepackage{amsmath,amssymb,amsfonts}
\usepackage{algorithm}
\usepackage{algpseudocode}
\usepackage{graphicx,color}
\usepackage{textcomp}
\usepackage{siunitx}
\usepackage{booktabs}
\def\BibTeX{{\rm B\kern-.05em{\sc i\kern-.025em b}\kern-.08em
    T\kern-.1667em\lower.7ex\hbox{E}\kern-.125emX}}

\usepackage{color,soul}
\usepackage{tikz}
\usetikzlibrary{shapes, arrows.meta, positioning, calc, decorations.pathreplacing, shadows}

\newcommand{\rev}[1]{\textcolor{black}{#1}}

\usepackage{url}

\begin{document}

\title{AI Lifecycle-Aware Feasibility Framework for Split-RIC Orchestration in NTN O-RAN}

\author{Daniele~Tarchi,~\IEEEmembership{Senior~Member,~IEEE}%

\thanks{D.~Tarchi is with the Department of Information Engineering, University of Florence, 50139 Firenze, Italy and with CNIT - University of Florence Research Unit, 50139 Firenze, Italy.}%
\thanks{This work was co-funded by the European Commission under the “5G-STARDUST” Project, which received funding from the Smart Networks and Services Joint Undertaking (SNS JU) under the European Union’s Horizon Europe research and innovation programme under Grant Agreement No 101096573.}%
\thanks{The views expressed are those of the authors and do not necessarily represent the project. The Commission is not liable for any use that may be made of any of the information contained therein.}%
\thanks{This work has received funding from the Swiss State Secretariat for Education, Research and Innovation (SERI).}%
\thanks{This work has been submitted to the IEEE for possible publication. Copyright may be transferred without notice, after which this version may no longer be accessible.}
}

\markboth{AI Lifecycle-Aware Feasibility Framework for Split-RIC Orchestration in NTN O-RAN}{Tarchi}

\maketitle

\begin{abstract}
Integrating Artificial Intelligence (AI) into Non-Terrestrial Networks (NTN) is constrained by the joint limits of satellite \rev{Size, Weight, and Power} (SWaP), feeder-link capacity, \rev{and orbital connectivity,} which directly \rev{affect} O-RAN closed-loop control and \rev{AI} lifecycle management. This paper studies the feasibility of distributing the O-RAN \rev{intelligence} across Ground, LEO, and GEO segments through \rev{three representative deployments}: (i) Ground-Centric control with telemetry streaming, (ii)~a ground--LEO Split-RIC with on-board inference and store-and-forward \rev{model refresh}, and (iii) GEO--LEO multi-layer control enabled by inter-satellite links. For each \rev{deployment}, we derive closed-form expressions for lifecycle energy, \rev{near-real-time control latency,} and \rev{model-refresh} latency\rev{, while explicitly enforcing power-feasible spaceborne computational throughput. Pairwise analytical boundaries and} numerical sensitivity analysis over \rev{telemetry volume,} feeder-link conditions, \rev{inference workload,} model \rev{longevity}, and orbital \rev{waiting time identify the operating regions in which different placements are energetically preferable or satisfy model-refresh requirements. The results show that the architecture choice is inherently lifecycle- and constraint-dependent, providing decision-enabling primitives for Service Management and Orchestration (SMO) level AI placement and lifecycle orchestration in NTN O-RAN.}
\end{abstract}

\begin{IEEEkeywords}
Service Management and Orchestration (SMO), O-RAN, RAN Intelligent Controller (RIC), Non-Terrestrial Networks (NTN), AI/ML lifecycle management, Edge Intelligence
\end{IEEEkeywords}

\section{Introduction}
\bstctlcite{IEEEexample:BSTcontrol}
The integration of Non-Terrestrial Networks (NTN) into the 6G ecosystem is poised to bridge the digital divide, providing ubiquitous coverage to underserved and remote regions \cite{9617565}. However, operating a multi-layer NTN that spans high-capacity Geostationary Earth Orbit (GEO) satellites and fast-moving Low Earth Orbit (LEO) constellations introduces time-varying topology, intermittent connectivity, and strong resource asymmetries. As a result, automation via Machine Learning (ML) is increasingly required to support scalable network and service management functions such as resource control, anomaly detection, and adaptive radio optimization~\cite{jsan13010014}. Recent research has therefore explored bringing computational capacity to on-board satellite nodes~\cite{shinde-tmlcn,10355087}. Yet, practical deployment of ML in space remains limited due to a fundamental paradox: modern Deep Learning (DL) models are computationally intensive, while satellite platforms are severely constrained by Size, Weight, and Power (SWaP). In particular, training and frequent model refresh operations may be incompatible with the energy budget of LEO payload computing without impacting mission operations.

Network softwarization through Virtual Network Functions (VNFs), Software Defined Networking (SDN), and network slicing has emerged as a key enabler to manage such heterogeneous infrastructures~\cite{10557759}. Within this context, the Open Radio Access Network (O-RAN) provides a reference architecture for virtualized and disaggregated RAN implementations through standardized interfaces (e.g., E2, A1, O1) and RAN Intelligent Controllers (RICs)~\cite{polese23_CSTO}. While O-RAN was originally designed for terrestrial deployments with abundant backhaul and grid power, its control-loop hierarchy is particularly attractive for NTN because it offers a natural separation between long-timescale functions (Non-Real-Time RIC) and latency-critical functions (Near-Real-Time RIC)~\cite{11101652}. In an NTN setting, this separation enables distributing intelligence across the ground--space boundary, but it also exposes a core service-management question: \emph{which parts of the AI lifecycle should be executed on-board versus offloaded, and at what management-plane cost over the feeder link?}

Despite this potential, the convergence of NTN, O-RAN, and AI is still hindered by two fundamental limitations in the state-of-the-art. First, many works treat space O-RAN as a direct mapping of terrestrial architectures to orbit~\cite{campana23, rihan23}, often assuming ideal connectivity or sufficient on-board compute. This leads to \emph{fragmented lifecycle modeling}: prior studies typically optimize a single phase of the AI pipeline (e.g., inference latency or distributed training convergence) but do not quantify the \emph{end-to-end} energy and time cost of the complete lifecycle, including training-data transfer, model dissemination, and repeated inference. In particular, the management-plane burden of transporting massive telemetry via O1 versus distributing compact model artifacts via A1 is rarely captured in a unified framework. Second, existing proposals lack \emph{feasibility boundaries}. They often advocate fixed ``all-edge'' or ``all-cloud'' deployments without identifying the physical regimes in which these choices are viable. Consequently, there is limited guidance that maps key NTN parameters, such as feeder-link conditions, orbital intermittency (waiting time), and model complexity, to the set of operationally feasible RIC placements.

In this paper, we address these gaps by investigating the feasibility of Split-RIC deployments tailored to SWaP-constrained NTN operation. We consider a decoupled AI lifecycle in which training and model refresh are \rev{assigned} to higher-capacity Ground or GEO nodes, while latency-\rev{sensitive} inference \rev{can be executed at} the LEO edge. Rather than proposing an online orchestration algorithm, \rev{our objective} is to \rev{derive} decision-enabling analytical primitives and operator-relevant feasibility regions that \rev{expose the energy, control-latency, and model-refresh trade-offs associated with different Ground/LEO/GEO placements}.

\rev{The resulting framework explicitly captures space-segment power constraints, communication overhead, feeder-link intermittency, and AI workload characteristics. The derived boundaries are intended to serve as constraints and performance indicators for future Service Management and Orchestration (SMO)-level placement, orchestration, and closed-loop assurance mechanisms rather than as fixed prescriptions for individual RIC applications.}
%
%
%
%
%The scope of this work is to provide a \emph{decision-enabling feasibility framework} for AI lifecycle management in NTN O-RAN, by deriving closed-form models and operational boundaries for distributing training, model updates, and near-real-time inference across Ground, LEO, and GEO segments. Our analysis quantifies how SWaP constraints, feeder-link conditions, and orbital intermittency jointly impact lifecycle energy and latency, and it delineates the physical regimes in which each Split-RIC deployment is viable. We aim to establish the analytical primitives and feasibility regions that can be used as constraints and performance indicators by future SMO-level optimization and closed-loop assurance solutions. We provide decision-enabling feasibility boundaries that can be directly used by the SMO to plan xApp/rApp placement and AI lifecycle operations under SWaP and feeder-link constraints. 
The specific contributions of this paper are as follows:
\begin{itemize}
    \item \textbf{Lifecycle-aware cost modeling for AI-enabled O-RAN management:}
    We derive closed-form expressions for \rev{space-segment} lifecycle energy and \rev{explicitly distinguish two operational} latency \rev{metrics: the latency of an individual near-real-time inference/control action and the model-refresh latency following an update trigger. The framework accounts for training-data transfer, model dissemination, inference inputs and outputs, repeated inference, and power-constrained spaceborne computation.}

    \item \textbf{Comparative feasibility analysis of Ground/LEO/GEO Split-RIC deployments:}
    We formalize and \rev{compare} three representative placements of O-RAN intelligence: Ground-Centric \rev{control} (S1), Ground--LEO Split-RIC with \rev{local} inference and intermittent \rev{Ground-based} learning (S2), and GEO--LEO multi-layer \rev{operation with GEO-based training over an inter-satellite link (ISL) (S3)}. \rev{Pairwise energy and model-refresh boundaries} quantify \rev{when local inference reduces communication cost and when} GEO \rev{assistance} can mitigate \rev{ground-contact waiting at the expense of additional space-segment energy}.

    \item \textbf{Operator-relevant feasibility regions and design guidelines:}
    \rev{Through} numerical sensitivity analysis over telemetry volume, feeder-link conditions, \rev{inference workload,} model \rev{longevity}, orbital waiting time, \rev{and model-update requirements,} we derive multi-dimensional feasibility regions \rev{for architecture selection}. The \rev{resulting maps distinguish energy preference from latency feasibility and} provide \rev{quantitative decision rules that can be incorporated into future SMO-level orchestration mechanisms}.

\end{itemize}

The remainder of this paper is organized as follows. Section~\ref{sec:Background} reviews O-RAN fundamentals, orbital NTN constraints, and related work on satellite edge intelligence from a management and lifecycle perspective. Section~\ref{sec:SystemModel} defines the system model by characterizing Ground/LEO/GEO resource profiles and formalizing the AI workload and its lifecycle cost components. Section~\ref{sec:Scenarios} introduces the three comparative deployments (Ground-Centric, Split-RIC, and Multi-Layer) and their management-plane data flows. Section~\ref{sec:Analysis} derives the analytical feasibility framework, \rev{including space-segment lifecycle-energy models, near-real-time control latency, model-refresh latency, and the corresponding pairwise feasibility boundaries}. Section~\ref{sec:Results} reports the numerical sensitivity analysis and resulting operational feasibility regions. Finally, Section~\ref{Sec:Con} concludes the paper and discusses future research directions.

\section{Background and Related Work}
\label{sec:Background}
This work addresses the engineering challenges inherent in the intersection of three distinct, yet interrelated, domains: the O-RAN software architecture, the physical constraints of orbital mechanics (specifically in LEO and GEO), and the computational costs of the AI lifecycle. To establish the necessary technical foundation for the proposed Split-RIC architecture, this section first deconstructs the core \mbox{O-RAN} components relevant to distributed intelligence. Then it characterizes the unique constraints of NTN and finally assesses the feasibility of integrating standard AI workflows into space. This analysis identifies the technical gaps in state-of-the-art satellite edge computing that our work aims to address.

\subsection{O-RAN Fundamentals for Distributed Intelligence}

The O-RAN architecture fundamentally disrupts the traditional RAN model by facilitating the disaggregation of the RAN into open, modular, and interoperable components, which were historically tightly coupled as monolithic \textit{black boxes}. For NTN and the proposed Split-RIC architecture, this modularity is not merely a convenience but a necessity for managing distributed satellite constellations. While the standard O-RAN specification is extensive \cite{polese23_CSTO}, we focus on the entities that are critical for the proposed architecture: the controllers (RICs) and the associated open interfaces (A1, E2 and O1). %In Fig. \ref{Fig:O_RAN}, the main elements of the O-RAN functional architecture are depicted.

%  \begin{figure}[tbp]
% \centering
% \includegraphics[width=0.9\columnwidth]{Images/Picture1.png}
% \caption{O-RAN Functional Architecture}
% \label{Fig:O_RAN} 
% \end{figure}

\subsubsection{The Intelligent Controllers (RICs) and Applications}
O-RAN introduces a hierarchical control loop mechanism driven by modular applications:
\begin{itemize}
    \item \textbf{Non-Real-Time RIC (Non-RT RIC):} Located within the SMO framework, this controller operates on a time scale $>\qty{1}{\second}$. It hosts rApps (i.e., Non-RT Applications), which are modular microservices responsible for high-level guidance, long-term analytics, and the training of heavy ML models using historical data. In \rev{the considered deployments}, rApps \rev{are hosted} on the Ground \rev{Segment} or\rev{, in S3, on a GEO node to exploit the larger computational resources available relative to the constrained LEO edge}.
    \item \textbf{Near-Real-Time RIC (Near-RT RIC):} This controller operates on a time scale between $\qty{10}{\milli\second}$ and $\qty{1}{\second}$. It hosts xApps (i.e., Near-RT Applications), which are lightweight, low-latency microservices designed to execute specific control logic (e.g., beam hopping, interference mitigation) and ML inference. In our proposed architecture, these xApps are deployed directly on the LEO satellite edge to ensure real-time responsiveness.
\end{itemize}

\subsubsection{Key Interfaces}
The disaggregation is enabled by standardized open interfaces:
\begin{itemize}
    \item \textbf{A1 Interface:} Connects the Non-RT RIC to the Near-RT RIC. Most importantly for this work, it can be extended to transport ML model weights~\cite{alliance2021ran}. In space, this maps to the Feeder Link, carrying model updates from Earth to Orbit.
    \item \textbf{E2 Interface:} Connects the Near-RT RIC to the E2 Nodes (O-CU and O-DU). It facilitates the collection of real-time telemetry and the enforcement of control actions directly on the satellite payload.
    \item \textbf{O1 Interface:} The management pipeline for FCAPS (Fault, Configuration, Accounting, Performance, Security). It allows the ground station to collect raw data traces. As we will discuss, overuse of this interface for data collection creates a bottleneck.
\end{itemize}

\subsection{Orbital Regimes and Constraints}
Deploying O-RAN software components in space, specifically within a NTN environment, is complicated by severe physical and channel constraints that differ drastically from terrestrial deployments. We must distinguish between the two primary orbital regimes, as they present opposite yet equally challenging constraints for AI and control loop deployment.

\subsubsection{Low Earth Orbit (LEO)}
LEO satellites operate in orbits close to Earth ($\approx\qtyrange{300}{2000}{\kilo\meter}$) and travel at high velocities ($\approx \qty{7.6}{\kilo\meter\per\second}$). Platforms in LEO (such as CubeSats) face strict SWaP constraints. Typical payload power budgets are below $\qty{100}{\watt}$, making it impractical to use power-intensive GPUs for on-board inference. In addition, LEO satellites have only short visibility windows with ground stations, %($\approx \qty{10}{\minute}$ per pass),
which makes transferring large volumes of data challenging. Although the LEO feeder-link Round-Trip Time (RTT) to a ground gateway can be comparatively small (typically on the order of 
$\approx \qtyrange{20}{40}{\milli\second}$ including propagation and protocol delays), it is still too high for latency-sensitive near-real-time control loops (e.g., handover or beam hopping) that often require response times below $\qty{10}{\milli\second}$ when the control loop is closed through the ground segment.

\subsubsection{Geostationary Orbit (GEO)}
GEO satellites remain essentially stationary at about $\qty{35786}{\kilo\meter}$ altitude. \rev{A single GEO--Ground hop introduces approximately 120~ms of one-way propagation delay at minimum slant range, while end-to-end GEO-relayed paths can incur several hundred milliseconds of round-trip delay.} %Signals require roughly $\qty{250}{\milli\second}$ to travel in each direction, giving a round-trip time exceeding $\qty{500}{\milli\second}$.
Although GEO platforms typically have kilowatt-scale power budgets, enough to support more capable onboard computing, the large latency makes real-time control from the ground infeasible. By the time a ground operator sends a command, the channel conditions have already shifted. Consequently, AI \textit{must} reside on-board to be effective.

\subsubsection{Feeder Link Constraints}
The wireless link between the satellite and the ground station, i.e., the Feeder Link, is a bottleneck. It carries aggregated user traffic, Telemetry, Tracking, and Command (TT\&C) data, and O-RAN control plane traffic. Transmitting raw I/Q samples for ground-based ML training consumes scarce bandwidth and high transmission power ($P_{tx}$), directly impacting the satellite's energy budget.

\subsection{The AI Lifecycle}
Implementing Machine Learning within a satellite network is not a singular event but a continuous lifecycle comprising data collection, model training, model deployment, and real-time inference. Unlike terrestrial networks where these phases often coexist in the same data center, the NTN environment necessitates a careful physical distribution of these tasks based on their specific computational and communication profiles.

\subsubsection{Training Phase}
The training of Deep Neural Networks (DNNs) is the most resource-intensive phase of the lifecycle. It involves iterative optimization algorithms (e.g., Stochastic Gradient Descent) and the calculation of gradients via back-propagation. \rev{Its computational workload and numerical precision depend on the specific model and learning procedure. In the deployments considered here, training is therefore assigned to higher-capacity Ground or GEO nodes, whereas the more frequent inference phase can be executed at the SWaP-constrained LEO edge.}%Training demands extensive matrix multiplications and, critically, high-precision floating-point formats (FP32 or FP64) to guarantee proper model convergence. In contrast, radiation-hardened satellite payloads are generally tailored for fixed-point arithmetic or low-bit-width integers (such as INT8) to minimize power consumption. Performing intensive floating-point computation directly on the spacecraft would generate excessive heat and drain the batteries quickly. Therefore, in our proposed architecture, the training process is fully centralized on high-capacity nodes—primarily within the Ground Segment, and potentially on GEO platforms. This approach exploits the substantially higher power budgets and High-Performance Computing (HPC) resources available in these environments relative to the highly constrained LEO edge.
 
\subsubsection{Inference Phase}
Inference represents the \textit{forward pass} of the neural network, used to generate control decisions based on live input data. While computationally cheaper than training, it is operationally demanding due to its high frequency (e.g., executing every Transmission Time Interval) and strict latency deadlines. In a traditional Bent-Pipe architecture, the satellite acts as a transparent relay, streaming raw telemetry (e.g., I/Q samples) to the ground for processing. While this spares satellite computing power, it saturates the scarce Feeder Link bandwidth and induces significant RTT latency, rendering real-time control infeasible. By moving the inference engine to the satellite edge, \rev{the high-volume input can be processed locally, avoiding its repeated feeder-link transmission at the cost of on-board computation energy.}
%we process raw data locally. This eliminates the need to transmit high-bandwidth telemetry, replacing it with negligible on-board energy consumption.

\subsubsection{The Fundamental Trade-offs}
The architectural challenge of Space O-RAN reduces to a multi-dimensional optimization problem involving two coupled trade-offs:
\begin{itemize}
    \item \textbf{Energy (Compute-vs-Communicate):} We must balance the \textit{Communication Energy} required to offload raw telemetry to the ground (high transmission power $P_{tx}$) against the \textit{Computation Energy} required to process that data locally (on-board battery drain).
    \item \textbf{Latency (Propagation-vs-Processing):} We must balance the \textit{Propagation Latency} inherent in ground-based loops (Round-Trip Time) against the \textit{Processing Latency} of constrained on-board accelerators. While ground servers are faster, they are physically distant; while satellites are close to the user, they are computationally slower.
\end{itemize}
Our feasibility analysis \rev{therefore separates energy preference from latency feasibility: local execution must satisfy the application-specific control deadline $T_{\mathrm{ctrl}}\leq\tau^{\max}$, while model-refresh requirements are evaluated separately through $T_{\mathrm{refresh}}$.}%aims to identify the operational \textit{Feasibility Region} where on-board inference minimizes total energy consumption while satisfying the stringent timing requirements ($\tau < \qty{10}{\milli\second}$) of the Near-RT RIC.

\subsection{Related Work and Gap Analysis}
\label{sec:RelatedWork}

The convergence of NTN, O-RAN architectures, and AI represents a nascent but rapidly evolving field. We categorize the state-of-the-art into three primary research streams.

\subsubsection{O-RAN Integration in NTN}
Early standardization efforts by 3GPP (Release 17/18) and the O-RAN Alliance have focused on defining the protocol stack adaptations required for NTN. Several studies have investigated the optimal placement of O-RAN functions (O-CU, O-DU, O-RU) in satellite payloads. Works such as \cite{campana23} and~\cite{rihan23} propose adaptive functional splits to handle the varying latency of LEO and GEO links. However, these contributions primarily address protocol synchronization and throughput optimization, treating the satellite as a static computing node rather than a dynamic energy-constrained platform. Research in \cite{11125758} and~\cite{10829759} explores hierarchical architectures where LEO satellites act as relays for GEO controllers. While these works define the topological connectivity, they often neglect the \textit{control plane overhead} introduced by the O-RAN interfaces (A1, E2, O1) over wireless feeder links. \rev{Recent architectural guidelines for intelligence deployment in NTN-based O-RAN further examine functional placement, control-loop timescales, and practical interface constraints \cite{TARCHI2027108723}. The present work complements this perspective by deriving quantitative lifecycle-energy and model-refresh feasibility boundaries.}

\subsubsection{AI and Machine Learning for Satellite Communications}
The application of ML to optimize satellite operations is well-documented, yet often decoupled from the O-RAN framework. Extensive literature exists on using Deep Reinforcement Learning (DRL) for beam hopping, power allocation, and interference management \cite{nguyen2024, qazzaz24}. However, the majority of these studies assume a ``Black Box" model deployment, where the energy cost of executing the inference engine is ignored~\cite{https://doi.org/10.1002/sat.1482}. Time-series forecasting models (e.g., Long Short-Term Memories (LSTMs), Transformers) have been proposed to predict traffic loads in LEO constellations \cite{10970731}. These works typically assume centralized training on the ground, without quantifying the bandwidth cost of continuously collecting high-resolution training data via the telemetry interface.

\subsubsection{Distributed Edge Intelligence and Lifecycle Management}
In the broader terrestrial 6G context, the concept of splitting AI training and inference is gaining attention. Techniques for distributing model training across edge nodes have been explored to preserve privacy and reduce bandwidth~\cite{8345716}. However, applying these terrestrial protocols to space encounters unique challenges, specifically, the intermittent connectivity of LEO satellites and the severe energy penalty of on-board back-propagation, which remains under-explored in current literature \cite{10694785}. While quantization and pruning are standard practices in embedded AI, their specific application to the O-RAN A1 interface for satellite links, where every bit of control overhead competes with user data, remains a gap. \rev{More directly related to the placement of computation in space, Denby and Lucia~\cite{Denby2020OEC} introduced the Orbital Edge Computing (OEC) paradigm, demonstrating how on-board processing and distributed nanosatellite pipelines can overcome the limitations of conventional bent-pipe architectures. Their framework explicitly considers orbital dynamics, energy availability, and ground-station access. Liu et al.~\cite{Liu2024Phoenix} subsequently proposed PHOENIX, a sunlight-aware scheduling framework that selects between in-orbit and Ground processing to reduce satellite battery consumption while meeting task-completion constraints. These contributions establish the relevance of energy-aware orbital computation and adaptive task offloading. However, they address different operational objectives from the present study, which focuses on the placement of O-RAN AI training and inference functions across Ground, LEO, and GEO nodes, explicitly accounting for model dissemination, repeated inference, and model-refresh deadlines.}

\subsubsection{Gap Analysis}
Despite these advancements, the current state of the art exhibits two fundamental limitations that prevent the practical deployment of O-RAN in space:
\begin{itemize}
    \item \rev{\textbf{Lifecycle-Aware Feasibility Modeling:} Existing satellite edge-computing studies have established the benefits of on-board processing and developed energy-aware task placement strategies. However, the joint cost of training-data transfer, model dissemination, and repeated near-real-time inference has received less attention in the specific context of O-RAN management functions distributed across NTN segments. In particular, the distinction between per-inference control latency and model-refresh latency is essential when evaluating architectural alternatives under intermittent connectivity.}
    \item \rev{\textbf{Operational Placement Boundaries:} Existing approaches address satellite task scheduling, functional split design, or distributed control using different objectives and resource assumptions. A complementary requirement is to identify explicit, parameter-dependent boundaries that determine when particular O-RAN AI placements minimize space-segment lifecycle energy while satisfying computation, control, and model-refresh constraints. Such boundaries can support SMO-level planning without requiring a particular online scheduling or orchestration algorithm.}
\end{itemize}

%\textbf{Our Contribution:} This work bridges these gaps by establishing the first rigorous Analytical Feasibility Framework for the Split-RIC architecture. Unlike previous qualitative surveys, we conduct a comprehensive sensitivity analysis to map the specific \textit{Feasibility Regions}, defined by link rate, data volume, and orbit, where On-Board AI is energetically and temporally superior to legacy terrestrial control.
\rev{\textbf{Our Contribution:}
Building on existing O-RAN NTN architectures and orbital edge-computing concepts, this work develops a lifecycle-aware analytical feasibility methodology for Ground/LEO/GEO placement of O-RAN intelligence. Rather than proposing a new task-scheduling algorithm or claiming novelty in on-board processing itself, we derive a unified space-segment energy model, enforce power-feasible computational throughput, and distinguish near-real-time control latency from model-refresh latency. The resulting pairwise analytical boundaries and multi-dimensional feasibility maps identify the conditions under which different deployments are energetically preferable or satisfy operational deadlines. These results provide complementary decision primitives for future SMO-level orchestration mechanisms.}

\section{System Model and Resource Formalization}\label{sec:SystemModel}
\rev{We consider a multi-layered NTN comprising Ground, LEO, and GEO nodes connected through potentially time-varying feeder links and ISLs.} In this dynamic topology, the set of nodes $\mathcal{V}$ is partitioned into two physically distinct domains: the resource-abundant Ground Segment ($\mathcal{G}_{st}$) and the resource-constrained Space Segment ($\mathcal{S}$). Unlike terrestrial networks where the RIC can be centralized in a metro-edge cloud with reliable power, the \textit{Space O-RAN} architecture must physically decouple the control hierarchy. We propose a Split-RIC framework where the learning lifecycle is distributed: the heavy training phases reside on Earth, or eventually on GEO satellites, while the latency-critical inference phases are pushed to the LEO satellite edge. To assess the feasibility of distributing AI workloads, we first formalize the physical domains (i.e., nodes), the connectivity layers (i.e., links), and the characteristics of the AI tasks.

\subsection{Physical Domains and Node Capabilities}
\label{subsec:PhyDom_NodeCap}
The network comprises three distinct physical domains. We characterize each node $n$ by the resource profile $\mathcal{R}_n = \langle F_n^{\rev{\mathrm{peak}}}, P_n^{\rev{\mathrm{AI}}}, \epsilon_n \rangle$, representing \rev{the nominal peak computational throughput, the power budget allocated to AI processing, and the hardware energy efficiency}, respectively. The adopted values, summarized in Table \ref{tab:NodeResources}, are derived from state-of-the-art hardware specifications for terrestrial and space-grade processors.

\rev{The peak throughput $F_n^{\mathrm{peak}}$ represents the nominal computational capability of the hardware and does not necessarily correspond to a continuously sustainable operating point. For the spaceborne nodes, the throughput available to AI workloads is constrained by the associated AI-processing power budget. We therefore define the power-feasible effective throughput as
\begin{equation}
F_n^{\mathrm{eff}}
=
\min\left(
F_n^{\mathrm{peak}},
\frac{P_n^{\mathrm{AI}}}{\epsilon_n}
\right),
\qquad
n\in\{\mathcal{S}_{LEO},\mathcal{S}_{GEO}\}.
\label{eq:effective_throughput}
\end{equation}
By construction, the resulting operating point satisfies
$\epsilon_n F_n^{\mathrm{eff}}\leq P_n^{\mathrm{AI}}$.
The effective throughput, rather than the nominal peak value, is therefore used when evaluating spaceborne computation latency.}

\begin{table}[tbp]
\caption{Node Resource Specifications}
\label{tab:NodeResources}
\centering
% This ensures the table fits exactly in the column width
\resizebox{\columnwidth}{!}{%
\begin{tabular}{@{}lccc@{}}
\toprule
\textbf{Param.} & \textbf{Ground ($\mathcal{G}_{st}$)} & \textbf{LEO ($\mathcal{S}_{LEO}$)} & \textbf{GEO ($\mathcal{S}_{GEO}$)} \\ 
\midrule
\textbf{HW Type} & Cloud GPU & Rad-Hard FPGA & Space DSP \\
\textbf{\rev{Peak Throughput ($F_n^{\mathrm{peak}}$)}} & $> \qty{1}{\peta\text{FLOP}\per\second}$ & $\approx \qty{10}{\tera\text{FLOP}\per\second}$ & $\approx \qty{200}{\tera\text{FLOP}\per\second}$ \\
\textbf{\rev{AI Power Budget ($P_n^{\mathrm{AI}}$)}} & \rev{Not constrained} & $\qty{20}{\watt}$ & $\qty{1000}{\watt}$ \\
\textbf{Eff. ($\epsilon_n$)} & $\qty{200}{\pico\joule}$/FLOP & $\qty{20}{\pico\joule}$/FLOP & $\qty{100}{\pico\joule}$/FLOP \\ 
\bottomrule
\end{tabular}%
}
\end{table}

\subsubsection{Ground Segment ($\mathcal{G}_{st}$)}
The Ground Segment $\mathcal{G}_{st}$ comprises the TT\&C ground stations and the associated terrestrial cloud infrastructure. It acts as the sink for global telemetry, capable of storing petabytes of historical dataset $\mathcal{D}$ without storage constraints and as the physical anchor for the NTN, providing backhaul connectivity and access to terrestrial data centers. Unlike space-based nodes, the ground segment is characterized by virtually unlimited power supply and High-Performance Computing clusters. We model the ground segment node $n_{gnd} \in \mathcal{G}_{st}$ as having \rev{non-binding} computational capacity relative to the space segment\rev{, formally represented as} $F_{\mathrm{GND}} \to \infty$ \rev{in the resource-feasibility model}. \rev{This assumption expresses that Ground computation is not a binding resource in the architectural feasibility comparison. For latency calculations, however, we retain a finite reference Ground throughput, denoted by $F_{\mathrm{GND}}$, whose numerical value is specified in Section~\ref{sec:Results}. Unlike the spaceborne nodes, $F_{\mathrm{GND}}$ is not derived from an AI power-budget constraint in the present model.} Although powerful, terrestrial GPUs (e.g., NVIDIA A100/H100) prioritize throughput over efficiency, resulting in a higher energy cost per operation ($\epsilon_{gnd} \approx \qty{200}{\pico\joule}/\text{FLOP}$) compared to specialized edge accelerators.

\subsubsection{Tier-1 Space Segment: LEO Edge ($\mathcal{S}_{LEO}$)}
LEO satellites ($\approx \qty{600}{\km}$) serve as the resource-constrained edge interface to user terminals. They are characterized by high orbital dynamics and severe SWaP constraints. The payload typically relies on energy-efficient hardware accelerators such as radiation-hardened Field Programmable Gate Arrays (FPGAs) or System-on-Chip (SoC) accelerators (e.g., Xilinx Versal AI Core, NVIDIA Jetson). The power budget available for \rev{AI} O-RAN processing, \rev{$P_{\mathrm{LEO}}^{\mathrm{AI}}$}, is strictly bounded by the solar array generation and battery depth (typically $\approx \qty{20}{\watt}$ for the computing payload). However, these devices achieve high efficiency ($\epsilon_{LEO} \approx \qty{20}{\pico\joule}/\text{FLOP}$) via low-precision integer arithmetic (INT8). \rev{Consequently, the computational throughput used in the latency analysis is $F_{\mathrm{LEO}}^{\mathrm{eff}}$, obtained from \eqref{eq:effective_throughput}, rather than the nominal peak throughput $F_{\mathrm{LEO}}^{\mathrm{peak}}$.} Due to their proximity to Earth, LEO nodes offer the lowest propagation delay to served users (single-hop delays on the order of a few to $\approx \qty{10}{\milli\second}$, depending on geometry). When inference and actuation are performed locally on-board (i.e., without traversing the feeder link), LEO nodes can therefore support near-real-time control loops.

\subsubsection{Tier-2 Space Segment: GEO Hubs ($\mathcal{S}_{GEO}$)}
Geostationary satellites ($\approx \qty{36000}{\km}$) act as high-capacity regional hubs. GEO platforms utilize larger buses capable of hosting specialized on-board processing payloads or powerful Digital Signal Processing (DSP) clusters. While not infinite like the ground, GEO power budgets are significantly higher ($\rev{P_{\mathrm{GEO}}^{\mathrm{AI}}} \approx \qty{1}{\kilo\watt}$), allowing for continuous operation of complex processors. \rev{The corresponding power-feasible computational throughput is denoted by $F_{\mathrm{GEO}}^{\mathrm{eff}}$ and is obtained from \eqref{eq:effective_throughput}.} The high propagation delay ($>\qty{250}{\milli\second}$) renders GEO nodes unsuitable for sub-millisecond radio loop control. However, their higher nominal compute capacity \rev{($F_{\mathrm{GEO}}^{\mathrm{peak}}
\gg F_{\mathrm{LEO}}^{\mathrm{peak}}$)} and extended visibility of compatible LEO satellites, subject to orbital geometry, make them suitable candidates for hosting aggregation tasks, distributed training, or latency-tolerant network orchestration.

\subsection{Interface Mapping and Link Dynamics}
The connectivity between the physical segments is governed by standard O-RAN logical interfaces. However, in an NTN environment, these logical interfaces must be mapped onto physical wireless links with distinct channel characteristics and resource constraints.

\subsubsection{Feeder Link (The Physical A1/O1 Interface)}
The Ground-to-Satellite Feeder Link serves as the physical transport layer for the vertical O-RAN interfaces. This link carries the A1 Interface (Policy/Model updates from Non-RT RIC to Near-RT RIC) and the O1 Interface (Management/Telemetry from Satellite to SMO). Unlike fiber-based terrestrial fronthaul, this link is a shared medium carrying User Plane traffic, TT\&C data, and O-RAN Control Plane messages. Bandwidth $B_{feed}$ is strictly limited and the achievable uplink data rate $R_{feed}(t)$ is governed by the Shannon limit, dependent on the time-varying slant range $d_{s,g}(t)$ and the satellite's transmission power $P_{tx}$. %Transmitting data from space to ground consumes significant power $P_{tx}^{feed}$. 
\rev{For the feeder link, $P_{\mathrm{tx,feed}}^{\mathrm{LEO}}$ and $P_{\mathrm{rx,feed}}^{\mathrm{LEO}}$ denote the LEO-side transmit and receive powers, respectively, while $R_{\mathrm{ul}}^{\mathrm{feed}}$ and $R_{\mathrm{dl}}^{\mathrm{feed}}$ denote the LEO-to-Ground and Ground-to-LEO data rates.}
\subsubsection{Inter-Satellite Link (Physical Xn/E2)} 
Optical inter-satellite links (OISLs) provide high-rate connectivity between satellites. These links carry the Xn Interface (Handover coordination between satellites) and, in distributed scenarios, the E2 Interface (e.g., if a Near-RT RIC hosted on one satellite controls an \rev{E2 node, such as an O-DU, hosted on another satellite}). \rev{In the baseline model, the GEO--LEO ISL provides a high-rate inter-satellite path characterized by $R_{\mathrm{ISL}}$ and by the corresponding node-side transmit and receive powers. The ISL is assumed to be available during the considered model-update interval, so that no additional ISL-access waiting time is included in the baseline S3 formulation. Intermittent ISL availability is considered separately in the robustness analysis. }%ISLs offer high bandwidth ($R_{ISL}$) and lower transmission energy per bit ($P_{tx}^{ISL}$) compared to RF feeder links. Connectivity is continuous between LEO and GEO, eliminating the ``waiting time" for ground station passes.
\rev{For the GEO--LEO ISL, $P_{\mathrm{tx,ISL}}^{n}$ and $P_{\mathrm{rx,ISL}}^{n}$ denote the transmit and receive powers of node $n\in\{\mathrm{LEO},\mathrm{GEO}\}$, $R_{\mathrm{ISL}}$ is the achievable ISL data rate, and $T_{\mathrm{prop,ISL}}$ is the one-way propagation delay.}

\subsection{AI Workload Characterization}
To facilitate the mapping of intelligence onto the physical nodes, we mathematically characterize the AI process. Unlike traditional task offloading models that consider a single computational block, we model the AI lifecycle as a coupled pair of distinct workloads: \textit{Inference} (executed frequently) and \textit{Training} (executed periodically). Let $\mathcal{K} = \{1, \dots, K\}$ denote the set of AI tasks required by the network. Each task $k \in \mathcal{K}$ is defined by a composite tuple $\Phi_k = \langle \mathcal{W}_k^{inf}, \mathcal{W}_k^{train}, \sigma_k \rangle$ corresponding to the Inference Profile, Training Profile and Model Artifact.

%The Inference Profile is a tuple $\mathcal{W}_k^{inf}=\langle \delta_k^{in}, \delta_k^{out}, \omega_k^{inf}, \tau_k^{max} \rangle$ that defines the near real-time execution requirements, where $\delta_k^{in}$ is the size of the raw input telemetry (e.g., one I/Q snapshot) required for an inference decision, $\delta_k^{out}$ is the size of the inference output (e.g., beam index control command), $\omega_k^{inf}$ is the computational complexity in FLOPs of the forward pass for the quantized model, and $\tau_k^{max}$ is the hard latency deadline (e.g., \qty{10}{\milli\second} for Near-RT RIC). Typically $\delta_k^{out} \ll \delta_k^{in}$, but it must be transported to the actuator (O-DU/O-RU).

\rev{The Inference Profile is defined as
\begin{equation}
\mathcal{W}^{\mathrm{inf}}_k = \left\langle \delta^{\mathrm{in}}_k, \delta^{\mathrm{out}}_k, \omega^{\mathrm{inf}}_k, \tau^{\max}_k \right\rangle ,
\end{equation}
where $\delta^{\mathrm{in}}_k$ is the size of the telemetry input required for one inference decision, $\delta^{\mathrm{out}}_k$ is the size of the resulting control output, $\omega^{\mathrm{inf}}_k$ is the computational workload of one forward pass, expressed in FLOPs, and $\tau^{\max}_k$ is the maximum admissible control-loop latency.} \rev{The output size $\delta^{\mathrm{out}}_k$ explicitly accounts for the control information generated by the inference process, e.g., a beam index, scheduling decision, or resource-allocation command. In the Ground-Centric scenario, this output must be transported from the Ground Near-RT RIC back to the satellite actuator and therefore contributes to both feeder-link energy and control-loop latency. In the Split-RIC scenarios, instead, inference and actuation are co-located on the LEO node and the corresponding E2 exchange is assumed to take place over the local on-board interconnect, whose communication cost is negligible compared with the external feeder and ISL.} \rev{The inference output is generally much smaller than the corresponding input telemetry, i.e., $\delta^{\mathrm{out}}\ll\delta^{\mathrm{in}}$. Nevertheless, it is explicitly retained in the analytical model whenever the control output must traverse an external link, as in the Ground-Centric scenario.}

The Training Profile is a tuple $\mathcal{W}_k^{train}=\langle \Delta_k^{train}, \Omega_k^{train}\rangle$ defining the model update requirements, occurring with a period $T_{update}$, where $\Delta_k^{train}$ is the size of the historical training dataset $\mathcal{D}$ and $\Omega_k^{train}$ is the computational complexity of the back-propagation process (epochs $\times$ batch size $\times$ gradient calculations). Note that $\Delta_k^{train} \gg \delta_k^{in}$, as it aggregates thousands of samples. The Model Artifact is $\sigma_k$, that corresponds to the size of the compressed model weights that must be transported from the Training Node to the Inference Node via the A1 interface.

%We map these tasks to the set of available physical nodes $\mathcal{V} = \{ \mathcal{G}_{st}, \mathcal{S}_{LEO}, \mathcal{S}_{GEO} \}$. Each node $n \in \mathcal{V}$ is characterized by its resource profile $\mathcal{R}_n = \langle F_n, P_n^{max}, \epsilon_n \rangle$ where $F_n$ is the maximum computational throughput of the node ($F_{Gnd} \gg F_{GEO} > F_{LEO}$), $P_n^{max}$  is he maximum power budget allocated for AI processing and $\epsilon_n$ is the energy efficiency of the hardware accelerator. Specialized LEO space-grade FPGAs typically offer superior energy efficiency (lower Joules/FLOP, i.e., $\epsilon_{LEO} < \epsilon_{Gnd}$), whereas terrestrial GPUs prioritize raw throughput ($F_{Gnd}$) at the cost of higher power consumption.

\subsection{Generalized Cost Model}
To evaluate different architectural splits agnostic of the specific scenario, we define a unified cost model. The total system cost is a function of two orthogonal metrics: the \textit{Energy Consumption} ($\mathcal{J}$) required to execute the task at the assigned node and to transmit the necessary data artifacts (input telemetry, training datasets, or model weights), and the \textit{End-to-End Latency} ($\mathcal{T}$) required to close the control loop or complete the learning cycle. Let $n_{proc}$ be the node assigned for processing (either inference or training) and $n_{src}$ be the node where the data originates.

\subsubsection{Energy Consumption Model ($\mathcal{J}$)}
The total energy cost is given by:
\begin{multline}
    \mathcal{J}_k(n_{src}, n_{proc}) =\\ E_{comp}(n_{proc}, \Omega) + E_{comm}(n_{src} \to n_{proc}, D)
\end{multline}
where $E_{comp}$ is the Computational Cost and $E_{comm}$ is the Communication Cost. The computational cost corresponds to the energy required to process a workload of complexity $\Omega$ on node $n_{proc}$:
    \begin{equation}
        E_{comp}(n_{proc}, \Omega) = \Omega \cdot \epsilon_{n_{proc}} 
        %+ P_{static} \cdot \frac{\Omega}{F_{n_{proc}}}
    \end{equation}
where $\epsilon_{n_{proc}}$ is the hardware efficiency (Joules/FLOP). The communication cost is the energy required to transmit a data volume $D$ from source $n_{src}$ to destination $n_{proc}$ over link $l$:
    \begin{multline}
        E_{comm}(n_{src} \to n_{proc}, D) =\\ 
        \begin{cases} 
            P_{tx}^{link} \cdot \frac{D}{R_{link}} & \text{if } n_{src} \neq n_{proc} \\
            0 & \text{if } n_{src} = n_{proc}
        \end{cases}
    \end{multline}
\rev{The expression above represents the transmitter-side contribution. In the space-segment lifecycle accounting of Section \ref{sec:Analysis}, the corresponding receiver-side energy is also included whenever reception occurs at a spaceborne node, while Ground-side energy consumption is excluded.}

\subsubsection{Latency Model ($\mathcal{T}$)}
The total latency is defined as the sum of the transmission delay, propagation delay, processing delay, and transmission window waiting delay:
\begin{multline}
    \mathcal{T}_k(n_{src}, n_{proc}) =\\ T_{comm}(n_{src} \to n_{proc}, D) + T_{comp}(n_{proc}, \Omega) + T_{wait}
\end{multline}
where $T_{comm}$ is the Communication Latency that includes serialization delay and physical propagation delay: 
    \begin{equation}
        T_{comm} = 
        \begin{cases} 
            \frac{D}{R_{link}} + \text{RTT}_{link}/2 & \text{if } n_{src} \neq n_{proc} \\
            0 & \text{if } n_{src} = n_{proc}
        \end{cases}
    \end{equation}
$T_{comp}$ is the Computation Latency corresponding to the time required to execute the workload on the target hardware:
    \begin{equation}
        T_{comp} = \frac{\Omega}{F_{n_{proc}}}
    \end{equation}
$T_{wait}$ corresponds to the amount of time that should be waited before the available transmission window (i.e., from/to a LEO satellite) occurs.

This generalized formulation allows us to quantify the cost of distinct architectural choices. For instance, when offloading to ground it incurs a high $T_{comm}$ (due to RTT) and potentially high $T_{wait}$, whereas for on-board processing it reduces $T_{comm}$ to zero but increases the on-board energy burden.

\subsection{Model Assumptions and Scope of Validity}\label{sec:Assumptions}
\rev{The baseline considers uncompressed telemetry and an available GEO--LEO ISL during each model-refresh operation. Training workload, transferred data, and model size are represented by aggregate quantities $\Omega^{train}$, $\Delta^{train}$, and $\sigma$, which may include multiple learning rounds. Spaceborne processing is limited by the power-feasible throughput defined in Section~\ref{subsec:PhyDom_NodeCap}, while application-specific inference-quality requirements must be satisfied independently. The sensitivity to these assumptions and the practical limitations of optical ISL deployment are discussed in Section~\ref{sec:Results}.}

\section{Comparative Deployment Architectures}
\label{sec:Scenarios}
Based on the system model defined above, we identify three distinct architectural configurations for the O-RAN control hierarchy. These scenarios, illustrated in Fig. \ref{fig:Scenarios}, represent specific mappings of the AI Workloads ($\mathcal{W}_{inf}, \mathcal{W}_{train}$) to the Physical Nodes ($\mathcal{G}_{st}, \mathcal{S}_{LEO}, \mathcal{S}_{GEO}$), creating distinct topology-dependent cost profiles.

\begin{figure*}[htbp]
    \centering
\resizebox{0.75\linewidth}{!}{%
\begin{tikzpicture}[
    % Global Styles
    font=\footnotesize\sffamily,
    >=Stealth,
    % Node Styles
    ground/.style={draw, fill=gray!20, minimum width=2cm, minimum height=0.3cm, anchor=south},
    gs_icon/.style={regular polygon, regular polygon sides=3, draw, fill=gray!80, minimum size=0.8cm, inner sep=0pt},
    sat_body/.style={rectangle, draw, fill=white, minimum width=0.8cm, minimum height=0.6cm, drop shadow},
    sat_panel/.style={rectangle, draw, fill=blue!30, minimum width=0.8cm, minimum height=0.4cm},
    % Function Icons
    brain/.style={cloud, cloud puffs=9.7, draw, fill=red!10, minimum width=0.8cm, aspect=1.5, inner sep=0pt},
    chip/.style={rectangle, draw, fill=green!20, minimum size=0.6cm},
    % Link Styles
    raw_link/.style={draw=red!80, line width=1.5mm, <->},
    model_link/.style={draw=green!60!black, line width=0.5mm, dashed, ->},
    isl_link/.style={draw=blue!80, line width=1mm, <->},
    inf_loop/.style={->, thick, black!70, bend right=90, looseness=2.5}
]

% =========================================================================
% MACRO: DRAW SATELLITE
% =========================================================================
\newcommand{\drawsat}[3]{ % #1=name, #2=x, #3=y, #4=label
    \node[sat_body] (#1) at (#2, #3) {}; 
    \node[sat_panel, anchor=east] at (#1.west) {};
    \node[sat_panel, anchor=west] at (#1.east) {};
    % Redraw body on top to hide panel overlap
    \node[sat_body] at (#2, #3) {\textbf{#1}}; 
}

% =========================================================================
% MACRO: DRAW GROUND STATION
% =========================================================================
\newcommand{\drawgs}[2]{ % #1=name, #2=position
    \node[gs_icon] (#1) at (#2) {};
    \draw[thick] (#1.north) -- ++(0, 0.3) arc(270:90:0.2); % Dish
    \node[right=0.1cm of #1] {Ground Station};
}

% =========================================================================
% PANEL A: SCENARIO 1 (GROUND CENTRIC)
% =========================================================================
\begin{scope}[shift={(0,0)}]
    % Earth
    \draw[thick, fill=gray!10] (-2.5, -0.5) rectangle (2.5, 0);
    \node at (0, -0.25) {EARTH};
    
    % Nodes
    \drawgs{GS1}{0, 0.225}
    \drawsat{LEO}{0}{4}
    
    % Internal Function Icons
    \node[brain, scale=0.6, anchor=center] at (GS1.center) {}; 
    \node[scale=0.6, text=red] at (0, -0.8) {Training + Inference};

    % Link
    \draw[raw_link] (GS1.north) -- (LEO.south) node[midway, right, align=left, black] {Feeder Link\\ \textbf{Raw Data}};
    
    % Scenario Label
    \node[draw, thick, fill=white, rounded corners] at (0, 5.5) {\textbf{(a) Scenario 1: Ground-Centric}};
\end{scope}

% =========================================================================
% PANEL B: SCENARIO 2 (SPLIT-RIC)
% =========================================================================
\begin{scope}[shift={(6,0)}]
    % Earth
    \draw[thick, fill=gray!10] (-2.5, -0.5) rectangle (2.5, 0);
    \node at (0, -0.25) {EARTH};
    
    % Nodes
    \drawgs{GS2}{0, 0.225}
    \drawsat{LEO}{0}{4}
    
    % Internal Function Icons
    \node[brain, scale=0.6] at (GS2.center) {}; % Training on Ground
    \node[chip, scale=0.5] at (LEO.south west) {}; % Inference on Sat (Chip icon)
    
    % Inference Loop
    \draw[inf_loop] (LEO.east) to (LEO.north);
    \node[right=0.5cm of LEO.north, scale=0.7] {Inference};

    % Link
    \draw[model_link] (GS2.north) -- (LEO.south) node[midway, right, align=left, black] {A1 Interface\\ \textbf{Model Weights}};
    
    % Waiting Time Icon (Clock)
    \draw[thick] (-1.3, 2) circle (0.25);
    \draw[thick] (-1.3, 2) -- ++(0, 0.15);
    \draw[thick] (-1.3, 2) -- ++(0.1, 0);
    \node[right] at (-1, 2) {$T_{wait}$};
    
    % Scenario Label
    \node[draw, thick, fill=white, rounded corners] at (0, 5.5) {\textbf{(b) Scenario 2: Split-RIC}};
\end{scope}

% =========================================================================
% PANEL C: SCENARIO 3 (GEO-LEO)
% =========================================================================
\begin{scope}[shift={(12,0)}]
    % Earth
    \draw[thick, fill=gray!10] (-2.5, -0.5) rectangle (2.5, 0);
    \node at (0, -0.25) {EARTH};
    
    % Nodes
    \drawgs{GS3}{0, 0.225} % Drawn but not connected
    \node[cross out, draw=red, thick, minimum size=0.5cm] at (GS3.center) {}; % Crossed out or inactive
    
    \drawsat{LEO}{0}{2.5}
    \drawsat{GEO}{0}{5} % Higher Up
    \node[above=0.1cm of GEO] {GEO Hub};

    % Internal Function Icons
    \node[brain, scale=0.6] at (GEO.center) {}; % Training on GEO
    \node[chip, scale=0.5] at (LEO.south west) {}; % Inference on LEO
    
    % Inference Loop
    \draw[inf_loop] (LEO.east) to (LEO.north);
    \node[right=0.5cm of LEO.north, scale=0.7] {Inference};

    % Link
    \draw[isl_link] (LEO.north) -- (GEO.south) node[midway, right, align=left, black] {ISL\\ \textbf{Training Data}};
    
    % Scenario Label
    \node[draw, thick, fill=white, rounded corners] at (0, 6.2) {\textbf{(c) Scenario 3: Multi-Layer}};
\end{scope}

\end{tikzpicture}}
    \caption{Comparative O-RAN Deployment Architectures. (a) S1: \rev{Ground-Centric processing with training and inference on the ground.} (b) S2: Proposed Split-RIC with inference at the edge and training on the ground. (c) S3: Hierarchical approach with inference at the edge and training on a GEO hub via ISL.}
    \label{fig:Scenarios}
\end{figure*}

\subsection{Scenario 1: Baseline Ground-Centric Control (S1)}
%This scenario represents the legacy ``Bent-Pipe" architecture (aligned with 3GPP Rel-17 NTN-TN interworking), where the satellite functions strictly as a transparent physical layer repeater.
\rev{Scenario 1 represents a Ground-Centric processing reference inspired by conventional transparent NTN operation, in which AI training and inference remain terrestrial while the LEO segment forwards telemetry and receives control information.} The entire O-RAN intelligence stack is centralized terrestrially. The Ground Segment hosts both the Non-RT RIC (for training) and the Near-RT RIC (for inference), treating the satellite node as a remote Radio Unit (O-RU) with no local decision-making capability. The operational data flow is structured in three phases: (i) the LEO satellite captures radio-frequency
signals, digitizes them, and forwards the assumed
uncompressed input telemetry ($\delta^{in}$)
to the Ground Segment over the feeder link; (ii) the terrestrial Near-RT RIC processes this massive data stream using high-performance GPUs to compute the necessary control action (e.g., beam weighting); %(iii) the control command is transmitted back to the satellite via the Feeder Link (mapped to the E2 interface) for actuation.
(iii) \rev{the resulting control action is returned over the feeder link for execution by the appropriate RAN function}. \rev{Because Ground computation lies outside the adopted space-segment energy boundary, S1 avoids spaceborne computation energy but incurs the repeated feeder-link cost associated with telemetry offloading and control-output reception.}

%While this architecture minimizes on-board SWaP costs ($E_{comp} \approx 0$), it places an immense burden on the feeder link. The system is strictly limited by the round-trip propagation delay ($\text{RTT}_{feed} \approx \qty{20}{\ms}$), rendering it infeasible for sub-millisecond control loops.

\rev{\subsection{Scenario 2: Ground--LEO Split-RIC (S2)}}

\rev{In Scenario~2, the AI lifecycle is split according to the
computational and connectivity characteristics of the two segments.
The Near-RT RIC and its xApps are hosted on the LEO node, whereas
the Non-RT RIC and model-training functions remain in the Ground
Segment. Consequently, raw telemetry is processed locally on-board
for each inference event, while historical samples are buffered and
periodically transferred to the Ground Segment for model retraining.}

\rev{The model-refresh procedure consists of three phases. First, the
LEO node waits until a suitable ground-contact opportunity and
uploads a training dataset of size $\Delta^{\mathrm{train}}$.
Second, the Ground Segment retrains the model with computational
workload $\Omega^{\mathrm{train}}$. Third, the resulting model
artifact of size $\sigma$ is returned to the LEO node through the
feeder link and activated by the on-board Near-RT RIC.}

\rev{We distinguish the waiting time before the training-data upload,
$T_{\mathrm{wait}}^{\mathrm{up}}$, from the additional waiting time
before the updated model can be returned,
$T_{\mathrm{wait}}^{\mathrm{ret}}$. This distinction avoids assuming
that two independent orbital waiting periods are always required.
In the baseline store-and-forward configuration, we conservatively
assume that the model is returned during a subsequent ground-contact
opportunity, such that
\begin{equation}
T_{\mathrm{wait}}^{\mathrm{up}}
=
T_{\mathrm{wait}}^{\mathrm{ret}}
=
T_{\mathrm{wait}}.
\end{equation}
A same-pass return, corresponding to
$T_{\mathrm{wait}}^{\mathrm{ret}}=0$, is considered separately in
the sensitivity analysis.}

\rev{Since the Near-RT RIC is co-located with the controlled O-DU/O-RU
functions on the LEO platform, inference outputs
$\delta^{\mathrm{out}}$ are exchanged locally through the on-board
interconnect. Their external communication cost is therefore
neglected in Scenario~2.}

\subsection{Scenario 3: Multi-Layer GEO-LEO Control (S3)}
This advanced scenario leverages the hierarchical nature of space networks to resolve the connectivity intermittency of S2. The control plane is fully non-terrestrial. The Near-RT RIC resides on the LEO Edge, while the Non-RT RIC is hosted on a high-capacity GEO Hub. Connectivity is maintained via ISL. The operational data flow is structured in three phases: (i) Similar to S2, real-time control is executed locally on the LEO node; \rev{(ii) upon a model-refresh trigger, the LEO satellite transfers the training data $\Delta^{train}$ to the GEO node via the ISL, assumed available during the considered update interval;}  (iii) The GEO node performs training and pushes updates back to LEO via the A1 interface over ISL. \rev{Under this baseline ISL-availability assumption, S3 avoids the ground-contact waiting terms affecting S1 and S2. This does not imply permanently available GEO--LEO connectivity or make S3 universally optimal: GEO-based training introduces additional computation energy and processing delay, so its main advantage arises when ground-contact intermittency would otherwise prevent timely model refresh. Intermittent ISL availability is considered separately in the robustness analysis.}

%This enables Continuous Learning ($T_{wait} \approx 0$) regardless of the satellite's position over oceans or remote areas. It trades the high energy cost of RF Feeder Links for the typically lower energy cost per bit of Optical ISLs, making it optimal for sparse ground networks. 
It is important to note that while the GEO relay introduces a propagation delay of approx. \qty{120}{\milli\second} \rev{one way in the reference geometry}, this link is utilized exclusively for the Non-RT RIC functions (A1 policy updates and O1 telemetry aggregation). \rev{These management-plane operations occur on timescales of seconds to minutes.} The latency-critical inference loop remains local on the LEO node, \rev{so the GEO propagation delay does not enter the near-real-time inference/control loop considered for S3.} \rev{The implementation of S3 presupposes compatible GEO--LEO ISL terminals on both spacecraft, potentially requiring dedicated space-pointing optical apertures in addition to the Earth-facing communication payload. Terminal mass, electrical power consumption, pointing/acquisition/tracking (PAT) requirements, and available simultaneous links therefore constitute additional deployment constraints that are not captured by the logical O-RAN functional mapping alone.}

\section{Analytical Feasibility Framework}
\label{sec:Analysis}
To assess the feasibility of the \rev{considered} architectures, we derive closed-form expressions for \rev{space-segment} lifecycle energy\rev{, near-real-time control latency,} and \rev{model-refresh} latency. These expressions \rev{specialize} the generalized cost models from Section \ref{sec:SystemModel} to the \rev{three} deployment \rev{scenarios introduced in Section \ref{sec:Scenarios}}.

\subsection{Lifecycle Definition}
We define the feasibility metric based on a complete Learning-to-Inference Cycle. A single cycle consists of three sequential phases:
\begin{enumerate}
\item \textbf{Training Phase:} Offloading a historical dataset of size $\Delta^{train} = |\mathcal{D}|$ (bits) from the edge to the training node.
\item \textbf{Update Phase:} Transporting the trained model weights of size $\sigma$ (bits) back to the inference node.
\item \textbf{Inference Phase:} Performing $N_{inf}$ inference operations over the model's lifespan. Each operation requires $\delta^{in}$ input bits and $\omega^{inf}$ FLOPs.\end{enumerate}

The parameter $N_{inf}$ represents the \textit{Model Longevity}—i.e., how many inference decisions can be made before concept drift necessitates a new training cycle.

\rev{\subsection{Scenario 1: Ground-Centric Control (S1)}}

\rev{In Scenario~1, both model training and inference are executed in the Ground Segment, while the LEO node acts as a remote radio endpoint. Accordingly, the space-segment cost is entirely associated with feeder-link communication. For each inference event, the LEO node uploads the input telemetry $\delta^{\mathrm{in}}$ to the Ground Near-RT RIC and subsequently receives the corresponding control output $\delta^{\mathrm{out}}$. In addition, a historical dataset of size $\Delta^{\mathrm{train}}$ is transferred to the Ground Segment when a model-refresh cycle is triggered.}

\subsubsection{\rev{Space-Segment Lifecycle Energy}}
\rev{Under the adopted space-segment energy-accounting boundary, the energy consumed by terrestrial computing and gateway equipment is not included. The LEO node expends energy to upload the training dataset, repeatedly transmit the inference input, and receive the corresponding control output. Hence,
\begin{multline}
%\begin{aligned}
J_{S1}^{\mathrm{space}}=P_{\mathrm{tx,feed}}^{\mathrm{LEO}}\frac{\Delta^{\mathrm{train}}}{R_{\mathrm{ul}}^{\mathrm{feed}}}\\
+N_{\mathrm{inf}}\left(P_{\mathrm{tx,feed}}^{\mathrm{LEO}}\frac{\delta^{\mathrm{in}}}{R_{\mathrm{ul}}^{\mathrm{feed}}}+P_{\mathrm{rx,feed}}^{\mathrm{LEO}}\frac{\delta^{\mathrm{out}}}{R_{\mathrm{dl}}^{\mathrm{feed}}}\right).
%\end{aligned}
\label{eq:energy_s1}
\end{multline}}
\rev{The exclusion of Ground-side computation from $J_{S1}^{\mathrm{space}}$ does not imply that terrestrial training and inference consume no energy; rather, such consumption lies outside the adopted accounting boundary, whose purpose is to quantify the energy burden imposed on SWaP-constrained spaceborne resources.}

\subsubsection{\rev{Near-Real-Time Control Latency}}
\rev{For each control action, the inference input must be transferred from LEO to the Ground Segment, processed by the Ground Near-RT RIC, and the resulting control output returned to the satellite. Accordingly,
\begin{equation}
T_{\mathrm{ctrl},S1}=\frac{\delta^{\mathrm{in}}}{R_{\mathrm{ul}}^{\mathrm{feed}}}+\frac{\omega^{\mathrm{inf}}}{F_{\mathrm{GND}}}+RTT_{\mathrm{feed}}+\frac{\delta^{\mathrm{out}}}{R_{\mathrm{dl}}^{\mathrm{feed}}}.
\label{eq:tctrl_s1}
\end{equation}}
\rev{The corresponding control-loop feasibility condition is
\begin{equation}
T_{\mathrm{ctrl},S1}\leq \tau^{\max}.
\label{eq:tctrl_feas_s1}
\end{equation}}
\rev{The waiting time associated with orbital visibility is not included in $T_{\mathrm{ctrl},S1}$, since this metric characterizes the latency of an individual control action while a feeder-link connection is available. Lack of feeder-link availability instead constitutes a service-availability constraint for Scenario~1.}

\subsubsection{\rev{Model-Refresh Latency}}
\rev{Since both model training and subsequent inference remain in the Ground Segment, the updated model does not need to be transmitted back to the LEO node. The model-refresh latency is therefore
\begin{equation}
T_{\mathrm{refresh},S1} = T_{\mathrm{wait}}^{\mathrm{up}}+\frac{\Delta^{\mathrm{train}}}{R_{\mathrm{ul}}^{\mathrm{feed}}}+\frac{RTT_{\mathrm{feed}}}{2}+\frac{\Omega^{\mathrm{train}}}{F_{\mathrm{GND}}}.
\label{eq:trefresh_s1}
\end{equation}
Here, $T_{\mathrm{wait}}^{\mathrm{up}}$ denotes the waiting time between the triggering of a model update and the first suitable ground-contact opportunity for uploading the training dataset. Once training is completed, the refreshed model is immediately available to the Ground Near-RT RIC, and no additional model-distribution delay is incurred.}

% \subsection{Scenario 1: Baseline Ground-Centric (S1)}
% In this baseline scenario, the satellite acts as a transparent relay. Both training and inference occur in the Ground Segment ($\mathcal{G}_{st}$).
% \subsubsection{Lifecycle Energy ($\mathcal{J}_{S1}$)}
% The satellite expends energy strictly on communication: transmitting the training dataset and, subsequently, streaming raw telemetry for every inference event:
% \begin{equation}
% \mathcal{J}_{S1} = \underbrace{P_{tx}^{feed} \cdot \frac{\Delta^{train}}{R_{ul}^{feed}}}_{\text{Training Offload}} + \underbrace{N_{inf} \cdot \left( P_{tx}^{feed} \cdot \frac{\delta^{in}}{R_{ul}^{feed}} \right)}_{\text{Real-Time Inference Streaming}}
% \end{equation}
% $\mathcal{J}_{S1}$ scales linearly with $N_{inf}$ and $\delta^{in}$. For data-intensive xApps (e.g., spectral analysis), the second term dominates, rapidly draining the LEO battery.

% \subsubsection{Lifecycle Latency ($\mathcal{T}_{S1}$)}
% The cycle latency includes the orbital waiting time to offload training data, plus the Round-Trip Time (RTT) for every inference action:
% \begin{equation}
% \mathcal{T}_{S1} = T_{wait}^{gnd} + \frac{\Delta^{train}}{R_{ul}^{feed}} + T_{train}^{gnd} + N_{inf} \cdot \left( \text{RTT}_{feed} + \frac{\delta^{in}}{R_{ul}^{feed}} \right)
% \end{equation}
% The inference loop suffers from the Feeder Link propagation delay ($\approx \qty{20}{\ms}$), rendering this architecture infeasible for control loops requiring $\tau < \qty{10}{\ms}$.

\rev{\subsection{Scenario 2: Ground--LEO Split-RIC (S2)}}

\rev{In Scenario~2, model training is executed in the Ground Segment, whereas inference is performed locally on the LEO node. We therefore distinguish the space-segment lifecycle energy, the near-real-time control latency, and the model-refresh latency.}

\subsubsection{\rev{Space-Segment Lifecycle Energy}}

\rev{The LEO node expends energy to upload the training dataset, receive the updated model artifact, and execute $N_{\mathrm{inf}}$ local inference operations. Hence,
\begin{multline}
%\begin{aligned}
J_{S2}^{\mathrm{space}}=\\ 
P_{\mathrm{tx,feed}}^{\mathrm{LEO}}\frac{\Delta^{\mathrm{train}}}     {R_{\mathrm{ul}}^{\mathrm{feed}}}+P_{\mathrm{rx,feed}}^{\mathrm{LEO}}\frac{\sigma}     {R_{\mathrm{dl}}^{\mathrm{feed}}} +N_{\mathrm{inf}}\omega^{\mathrm{inf}}\epsilon_{\mathrm{LEO}} .
%\end{aligned}
\label{eq:energy_s2}
\end{multline}}
\rev{Ground-side training energy is not included in $J_{S2}^{\mathrm{space}}$, since the adopted energy-accounting boundary includes only energy consumed by spaceborne resources. This does not imply zero physical energy consumption at the Ground Segment, but rather reflects the focus of the analysis on SWaP-constrained satellite resources.}

\subsubsection{\rev{Near-Real-Time Control Latency}}
\rev{Since inference and actuation are both performed on-board, the external feeder link is not traversed during an individual near-real-time control action. Neglecting the comparatively small latency of the local on-board E2 exchange, the control-loop latency is therefore
\begin{equation}
T_{\mathrm{ctrl},S2} = \frac{\omega^{\mathrm{inf}}}     {F_{\mathrm{LEO}}^{\mathrm{eff}}}.
\label{eq:tctrl_s2}
\end{equation}}
\rev{The corresponding feasibility condition for an xApp with latency requirement $\tau^{\max}$ is
\begin{equation}
\frac{\omega^{\mathrm{inf}}}{F_{\mathrm{LEO}}^{\mathrm{eff}}}\leq \tau^{\max}.
\label{eq:tctrl_feas_s2}
\end{equation}}
\rev{Using the power-feasible throughput $F_{\mathrm{LEO}}^{\mathrm{eff}}=1$~TFLOP/s adopted in the numerical evaluation, a workload of $w$~GFLOP per inference requires approximately $w$~ms of on-board processing time.}

\subsubsection{\rev{Model-Refresh Latency}}
\rev{The model-refresh latency measures the elapsed time from the triggering of a model update until the refreshed model becomes available at the LEO inference node. It is given by
\begin{multline}
%\begin{aligned}
T_{\mathrm{refresh},S2}=T_{\mathrm{wait}}^{\mathrm{up}}+\frac{\Delta^{\mathrm{train}}}{R_{\mathrm{ul}}^{\mathrm{feed}}}+\frac{RTT_{\mathrm{feed}}}{2}\\
+\frac{\Omega^{\mathrm{train}}}{F_{\mathrm{GND}}}+T_{\mathrm{wait}}^{\mathrm{ret}}+\frac{\sigma}{R_{\mathrm{dl}}^{\mathrm{feed}}}+\frac{RTT_{\mathrm{feed}}}{2}.
%\end{aligned}
\label{eq:trefresh_s2}
\end{multline}
Equivalently,
\begin{multline}
%\begin{aligned}
T_{\mathrm{refresh},S2}=T_{\mathrm{wait}}^{\mathrm{up}}+T_{\mathrm{wait}}^{\mathrm{ret}}+\frac{\Delta^{\mathrm{train}}}{R_{\mathrm{ul}}^{\mathrm{feed}}}\\
+\frac{\Omega^{\mathrm{train}}}{F_{\mathrm{GND}}}+\frac{\sigma}{R_{\mathrm{dl}}^{\mathrm{feed}}}+RTT_{\mathrm{feed}} .
%\end{aligned}
\label{eq:trefresh_s2_compact}
\end{multline}
For the baseline store-and-forward assumption, $T_{\mathrm{wait}}^{\mathrm{up}}=T_{\mathrm{wait}}^{\mathrm{ret}}=T_{\mathrm{wait}}$, yielding
\begin{equation}
T_{\mathrm{refresh},S2}^{\mathrm{baseline}}=2T_{\mathrm{wait}}+\frac{\Delta^{\mathrm{train}}}{R_{\mathrm{ul}}^{\mathrm{feed}}}+\frac{\Omega^{\mathrm{train}}}{F_{\mathrm{GND}}}+\frac{\sigma}{R_{\mathrm{dl}}^{\mathrm{feed}}}+RTT_{\mathrm{feed}} .
\label{eq:trefresh_s2_baseline}
\end{equation}
%
% With the reference parameters $\Delta^{\mathrm{train}}=10$~Gbit, $\sigma=50$~Mbit, $R_{\mathrm{ul}}^{\mathrm{feed}}
% =R_{\mathrm{dl}}^{\mathrm{feed}}=500$~Mbit/s, $RTT_{\mathrm{feed}}=20$~ms, $\Omega^{\mathrm{train}}=3$~PFLOP, and $F_{\mathrm{GND}}=1$~PFLOP/s, this becomes
% \begin{equation}
% %\boxed{
% T_{\mathrm{refresh},S2}^{\mathrm{baseline}}=2T_{\mathrm{wait}}+23.12~\mathrm{s}.
% %}
% \label{eq:trefresh_s2_numeric}
% \end{equation}
}
%
% \subsection{Scenario 2: Proposed Split-RIC (S2)}
% In the Split-RIC architecture, training remains on the Ground, but inference is moved to the LEO Edge ($\mathcal{S}_{LEO}$).
% \subsubsection{Lifecycle Energy ($\mathcal{J}_{S2}$)}
% The satellite incurs a one-time transmission cost for the training data and a reception cost for the model weights. Notably, the inference energy is now governed by the on-board hardware efficiency ($\epsilon_{LEO}$) rather than the radio link:
% \begin{multline}
% \mathcal{J}_{S2} =\\
% \underbrace{P_{tx}^{feed} \cdot \frac{\Delta^{train}}{R_{ul}^{feed}}}_{\text{Training Offload}} + \underbrace{P_{rx}^{feed} \cdot \frac{\sigma}{R_{dl}^{feed}}}_{\text{Model Rx}} + \underbrace{N_{inf} \cdot (\omega^{inf} \cdot \epsilon_{LEO})}_{\text{On-Board Inference}}
% \end{multline}
% Since $\omega^{inf} \cdot \epsilon_{LEO} \ll P_{tx}^{feed} \cdot \frac{\delta^{in}}{R_{ul}}$, this architecture becomes increasingly energy-efficient as $N_{inf}$ grows. This represents the Compute-vs-Communicate Gain.
% \subsubsection{Lifecycle Latency ($\mathcal{T}_{S2}$)}
% While inference is instantaneous ($T_{comp} \approx 0$), the \textit{Learning Latency} is penalized by the intermittency of the ground link. The satellite must wait for a pass to upload data, and potentially wait for a subsequent pass to download the model:
% \begin{equation}
% \mathcal{T}_{S2} = \underbrace{2 \cdot T_{wait}^{gnd}}_{\text{Double Wait Penalty}} + \frac{\Delta^{train}}{R_{ul}^{feed}} + T_{train}^{gnd} + N_{inf} \cdot T_{comp}^{LEO}
% \end{equation}
%
\subsection{\rev{Scenario 3: GEO--LEO Multi-Layer Control (S3)}}

\rev{In Scenario~3, the AI lifecycle is entirely supported by spaceborne resources. The Near-RT RIC and its xApps remain on the LEO node, whereas the Non-RT RIC and model-training functions are hosted by the GEO node. The LEO satellite transfers the training dataset to GEO through the ISL, and the updated model artifact is subsequently returned to LEO over the same link.}

\subsubsection{\rev{Space-Segment Lifecycle Energy}}
\rev{Unlike Scenarios~1 and~2, both endpoints involved in model retraining belong to the space segment. The lifecycle energy must therefore include the LEO and GEO communication costs, GEO training energy, and repeated LEO inference energy:
\begin{multline}
%\begin{aligned}
J_{S3}^{\mathrm{space}}=P_{\mathrm{tx,ISL}}^{\mathrm{LEO}}\frac{\Delta^{\mathrm{train}}}{R_{\mathrm{ISL}}}+P_{\mathrm{rx,ISL}}^{\mathrm{GEO}}\frac{\Delta^{\mathrm{train}}}{R_{\mathrm{ISL}}}+\Omega^{\mathrm{train}}\epsilon_{\mathrm{GEO}}\\
+P_{\mathrm{tx,ISL}}^{\mathrm{GEO}}\frac{\sigma}{R_{\mathrm{ISL}}}+P_{\mathrm{rx,ISL}}^{\mathrm{LEO}}\frac{\sigma} {R_{\mathrm{ISL}}}
+N_{\mathrm{inf}}\omega^{\mathrm{inf}}\epsilon_{\mathrm{LEO}} .
%\end{aligned}
\label{eq:energy_s3}
\end{multline}
When identical transmit and receive power values are assumed for the LEO and GEO ISL terminals, i.e., $P_{\mathrm{tx,ISL}}^{\mathrm{LEO}} =P_{\mathrm{tx,ISL}}^{\mathrm{GEO}} =P_{\mathrm{tx,ISL}}$ and $P_{\mathrm{rx,ISL}}^{\mathrm{LEO}} =P_{\mathrm{rx,ISL}}^{\mathrm{GEO}} =P_{\mathrm{rx,ISL}}$, \eqref{eq:energy_s3} reduces to
\begin{multline}
J_{S3}^{\mathrm{space}} = \left(P_{\mathrm{tx,ISL}}+P_{\mathrm{rx,ISL}}\right)\frac{\Delta^{\mathrm{train}}+\sigma}{R_{\mathrm{ISL}}}\\
+\Omega^{\mathrm{train}}\epsilon_{\mathrm{GEO}}+N_{\mathrm{inf}}\omega^{\mathrm{inf}}\epsilon_{\mathrm{LEO}} .
\label{eq:energy_s3_compact}
\end{multline}}

\subsubsection{\rev{Near-Real-Time Control Latency}}
\rev{As in Scenario~2, inference and actuation are co-located on the LEO node. The GEO segment participates only in the model-refresh process and is therefore not traversed during an individual near-real-time control action. Neglecting the comparatively small latency of the local on-board E2 exchange, 
\begin{equation}
T_{\mathrm{ctrl},S3} = \frac{\omega^{\mathrm{inf}}}{F_{\mathrm{LEO}}^{\mathrm{eff}}}.
\label{eq:tctrl_s3}
\end{equation}}
\rev{The corresponding control-loop feasibility condition is
\begin{equation}
\frac{\omega^{\mathrm{inf}}}{F_{\mathrm{LEO}}^{\mathrm{eff}}}\leq \tau^{\max}.
\label{eq:tctrl_feas_s3}
\end{equation}}
\rev{Consequently, $T_{\mathrm{ctrl},S3}=T_{\mathrm{ctrl},S2}$ under the adopted local-inference model; the distinction between the two scenarios arises from the placement and connectivity of the model-training function rather than from the near-real-time inference loop.}

\subsubsection{\rev{Model-Refresh Latency}}
\rev{Scenario~3 avoids the orbital waiting times associated with terrestrial ground-contact opportunities. A model-refresh cycle requires transferring the training dataset from LEO to GEO, executing the training workload on the GEO node, and returning the updated model to LEO. Hence,
\begin{equation}
%\begin{aligned}
T_{\mathrm{refresh},S3}=\frac{\Delta^{\mathrm{train}}}{R_{\mathrm{ISL}}}+T_{\mathrm{prop,ISL}}+\frac{\Omega^{\mathrm{train}}}{F_{\mathrm{GEO}}^{\mathrm{eff}}}\\
+\frac{\sigma}{R_{\mathrm{ISL}}}+T_{\mathrm{prop,ISL}},
%\end{aligned}
\label{eq:trefresh_s3}
\end{equation}
where $T_{\mathrm{prop,ISL}}$ denotes the one-way LEO--GEO propagation delay. Defining
\begin{equation}
RTT_{\mathrm{ISL}} = 2T_{\mathrm{prop,ISL}},
\label{eq:rtt_isl}
\end{equation}
the model-refresh latency can equivalently be written as
\begin{equation}
T_{\mathrm{refresh},S3} = \frac{\Delta^{\mathrm{train}}}{R_{\mathrm{ISL}}}+\frac{\Omega^{\mathrm{train}}}{F_{\mathrm{GEO}}^{\mathrm{eff}}}+\frac{\sigma}{R_{\mathrm{ISL}}}+RTT_{\mathrm{ISL}}.
\label{eq:trefresh_s3_compact}
\end{equation}}

\rev{This expression explicitly separates communication serialization, GEO processing, and propagation delay. In contrast to Scenario~2, no ground-contact waiting term is present under the baseline assumption that the GEO--LEO ISL remains available during the model-refresh operation.}

\subsection{Operational Feasibility Boundaries}
\label{sec:FeasibilityConditions}
\rev{The metrics introduced above characterize two distinct aspects of the AI lifecycle. Space-segment lifecycle energy quantifies the energy burden imposed on the spaceborne infrastructure, whereas $T_{\mathrm{ctrl}}$ and $T_{\mathrm{refresh}}$ capture, respectively, the latency of an individual near-real-time control action and the latency required to complete a model-refresh operation. Therefore, rather than defining a single global latency metric, we derive separate pairwise energy-preference boundaries, control-loop feasibility conditions, and model-refresh latency boundaries for the three deployment scenarios.}

%To determine the global optimal strategy, we generalize the comparison to identify the specific dominance regions for all three architectures. We define the feasibility boundaries for Energy ($\mathcal{J}$) and Latency ($\mathcal{T}$) separately.
{\color{black}

\subsubsection{Space-Segment Energy Preference}

The energy-preferred architecture is the one minimizing $J_{S_s}^{\mathrm{space}}$. Using
\eqref{eq:energy_s1}, \eqref{eq:energy_s2}, and \eqref{eq:energy_s3_compact}, the exact pairwise boundaries are obtained by equating the corresponding lifecycle energies.

For S1 and S2, the common training-data upload term cancels, yielding
\begin{subequations}
\label{eq:energy_boundaries}
\begin{align}
J_{S1}^{\mathrm{space}} = J_{S2}^{\mathrm{space}} \Longleftrightarrow & P_{\mathrm{tx,feed}}^{\mathrm{LEO}} \frac{\delta^{\mathrm{in}}}{R_{\mathrm{ul}}^{\mathrm{feed}}} + P_{\mathrm{rx,feed}}^{\mathrm{LEO}} \frac{\delta^{\mathrm{out}}}{R_{\mathrm{dl}}^{\mathrm{feed}}}\nonumber\\
&=\omega^{\mathrm{inf}}\epsilon_{\mathrm{LEO}}+\frac{P_{\mathrm{rx,feed}}^{\mathrm{LEO}}\sigma}{N_{\mathrm{inf}}R_{\mathrm{dl}}^{\mathrm{feed}}}.
\label{eq:energy_boundary_s1_s2}
\end{align}

For S2 and S3, the repeated LEO inference-energy term is common
to both architectures and therefore cancels:
\begin{align}
J_{S2}^{\mathrm{space}} = J_{S3}^{\mathrm{space}} \Longleftrightarrow & P_{\mathrm{tx,feed}}^{\mathrm{LEO}} \frac{\Delta^{\mathrm{train}}}{R_{\mathrm{ul}}^{\mathrm{feed}}}+P_{\mathrm{rx,feed}}^{\mathrm{LEO}}\frac{\sigma}{R_{\mathrm{dl}}^{\mathrm{feed}}}\nonumber\\
&=\left(P_{\mathrm{tx,ISL}}+P_{\mathrm{rx,ISL}}\right)\frac{\Delta^{\mathrm{train}}+\sigma}{R_{\mathrm{ISL}}}\nonumber\\
&\quad+\Omega^{\mathrm{train}}\epsilon_{\mathrm{GEO}} .
\label{eq:energy_boundary_s2_s3}
\end{align}

Finally, the S1--S3 boundary is
\begin{align}
J_{S1}^{\mathrm{space}}=J_{S3}^{\mathrm{space}}\Longleftrightarrow&P_{\mathrm{tx,feed}}^{\mathrm{LEO}}\frac{\Delta^{\mathrm{train}}}{R_{\mathrm{ul}}^{\mathrm{feed}}}\nonumber\\
&\hspace{-3.32051pt}+N_{\mathrm{inf}}\left(P_{\mathrm{tx,feed}}^{\mathrm{LEO}}\frac{\delta^{\mathrm{in}}}{R_{\mathrm{ul}}^{\mathrm{feed}}}+P_{\mathrm{rx,feed}}^{\mathrm{LEO}}\frac{\delta^{\mathrm{out}}}{R_{\mathrm{dl}}^{\mathrm{feed}}}\right)\nonumber\\
&=\left(P_{\mathrm{tx,ISL}}+P_{\mathrm{rx,ISL}}\right)\frac{\Delta^{\mathrm{train}}+\sigma}    {R_{\mathrm{ISL}}}\nonumber\\
&\quad+\Omega^{\mathrm{train}}\epsilon_{\mathrm{GEO}}+N_{\mathrm{inf}}\omega^{\mathrm{inf}}\epsilon_{\mathrm{LEO}} .
\label{eq:energy_boundary_s1_s3}
\end{align}
\end{subequations}

For each pair, the scenario whose corresponding side of~\eqref{eq:energy_boundaries} is smaller is energetically preferred. Equation~\eqref{eq:energy_boundary_s1_s2} also shows that the commonly used compute-versus-communicate condition is recovered only asymptotically when the one-time model-distribution cost is amortized over a large $N_{\mathrm{inf}}$ and $\delta^{\mathrm{out}}\ll\delta^{\mathrm{in}}$.

\subsubsection{Near-Real-Time Control Feasibility}
Control-loop feasibility is evaluated independently of model-refresh latency. Scenario~1 satisfies the xApp control deadline $\tau^{\max}$ when
\begin{subequations}
\label{eq:control_boundaries}
\begin{equation}
\frac{\delta^{\mathrm{in}}}{R_{\mathrm{ul}}^{\mathrm{feed}}}+\frac{\omega^{\mathrm{inf}}}{F_{\mathrm{GND}}}+\frac{\delta^{\mathrm{out}}}{R_{\mathrm{dl}}^{\mathrm{feed}}}+RTT_{\mathrm{feed}}\leq\tau^{\max}.
\label{eq:control_feas_s1}
\end{equation}

For Scenarios~2 and~3, inference and actuation are local to the LEO node, and the feasibility condition becomes
\begin{equation}
\frac{\omega^{\mathrm{inf}}}{F_{\mathrm{LEO}}^{\mathrm{eff}}}\leq\tau^{\max}.
\label{eq:control_feas_s2_s3}
\end{equation}

The pairwise control-latency crossover between S1 and the local-inference architectures is therefore
\begin{equation}
\frac{\delta^{\mathrm{in}}}{R_{\mathrm{ul}}^{\mathrm{feed}}}+\frac{\omega^{\mathrm{inf}}}{F_{\mathrm{GND}}}+\frac{\delta^{\mathrm{out}}}     {R_{\mathrm{dl}}^{\mathrm{feed}}}+RTT_{\mathrm{feed}}=\frac{\omega^{\mathrm{inf}}}{F_{\mathrm{LEO}}^{\mathrm{eff}}}.
\label{eq:control_boundary_s1_s2s3}
\end{equation}
\end{subequations}

Since $T_{\mathrm{ctrl},S2}=T_{\mathrm{ctrl},S3}$ under the adopted local-inference model, no S2--S3 crossover exists for the near-real-time control latency. Their distinction instead arises from the model-refresh process.

\subsubsection{Model-Refresh Latency Preference}

The pairwise model-refresh boundaries follow from \eqref{eq:trefresh_s1}, \eqref{eq:trefresh_s2_compact}, and \eqref{eq:trefresh_s3_compact}.

For S1 and S2, assuming the same initial waiting time $T_{\mathrm{wait}}^{\mathrm{up}}$, the difference is
\begin{subequations}
\label{eq:refresh_boundaries}
\begin{equation}
T_{\mathrm{refresh},S2}-T_{\mathrm{refresh},S1}=T_{\mathrm{wait}}^{\mathrm{ret}}+\frac{\sigma}{R_{\mathrm{dl}}^{\mathrm{feed}}}+\frac{RTT_{\mathrm{feed}}}{2}.
\label{eq:refresh_boundary_s1_s2}
\end{equation}
Since all terms on the right-hand side are non-negative, S1 always completes the model-refresh process no later than S2 under the adopted definitions. This follows because the refreshed model is already available at the Ground inference node in S1, whereas S2 must additionally return the model artifact to LEO.

The S1--S3 boundary is obtained by solving $T_{\mathrm{refresh},S1}=T_{\mathrm{refresh},S3}$ for the initial Ground waiting time:
\begin{multline}
T_{\mathrm{wait,BE}}^{\mathrm{up},S1-S3}=\frac{\Delta^{\mathrm{train}}}{R_{\mathrm{ISL}}}+\frac{\Omega^{\mathrm{train}}}{F_{\mathrm{GEO}}^{\mathrm{eff}}}+\frac{\sigma}{R_{\mathrm{ISL}}}+RTT_{\mathrm{ISL}}\\
-\left(\frac{\Delta^{\mathrm{train}}}{R_{\mathrm{ul}}^{\mathrm{feed}}}+\frac{RTT_{\mathrm{feed}}}{2}+\frac{\Omega^{\mathrm{train}}}{F_{\mathrm{GND}}}\right).
\label{eq:refresh_boundary_s1_s3}
\end{multline}
Thus, S3 provides a shorter model-refresh latency than S1 when $T_{\mathrm{wait}}^{\mathrm{up}}$ exceeds the threshold in
\eqref{eq:refresh_boundary_s1_s3}.

Finally, the S2--S3 boundary is
\begin{multline}
T_{\mathrm{wait}}^{\mathrm{up}}+T_{\mathrm{wait}}^{\mathrm{ret}}=\frac{\Delta^{\mathrm{train}}}{R_{\mathrm{ISL}}}+\frac{\Omega^{\mathrm{train}}}{F_{\mathrm{GEO}}^{\mathrm{eff}}}+\frac{\sigma}{R_{\mathrm{ISL}}}+RTT_{\mathrm{ISL}}\\
-\left(\frac{\Delta^{\mathrm{train}}}{R_{\mathrm{ul}}^{\mathrm{feed}}}+\frac{\Omega^{\mathrm{train}}}{F_{\mathrm{GND}}}+\frac{\sigma}{R_{\mathrm{dl}}^{\mathrm{feed}}}+RTT_{\mathrm{feed}}\right).
\label{eq:refresh_boundary_s2_s3}
\end{multline}
\end{subequations}
Accordingly, S3 provides the lower model-refresh latency when the total Ground-contact waiting time $T_{\mathrm{wait}}^{\mathrm{up}}+T_{\mathrm{wait}}^{\mathrm{ret}}$ exceeds the right-hand side of \eqref{eq:refresh_boundary_s2_s3}. Under the baseline store-and-forward assumption $T_{\mathrm{wait}}^{\mathrm{up}}=T_{\mathrm{wait}}^{\mathrm{ret}}=T_{\mathrm{wait}}$, the corresponding threshold becomes one half of this quantity. In the same-pass case, $T_{\mathrm{wait}}^{\mathrm{ret}}=0$, the full threshold applies to $T_{\mathrm{wait}}^{\mathrm{up}}$.}

\section{Performance Evaluation and Numerical Results}
\label{sec:Results}
% In this section, we numerically evaluate the analytical framework derived in Section \ref{sec:Analysis}. We perform a multi-dimensional sensitivity analysis to characterize the operational regions where each architecture (S1, S2, S3) is physically superior.  The performance of the proposed Split-RIC architectures (i.e., S2 and S3) is benchmarked against the State-of-the-Art (SoA) baseline, denoted as S1. This baseline corresponds to the conventional “bent-pipe” or transparent architecture used in legacy systems, where satellites serve solely as relay nodes and all control intelligence is concentrated in the terrestrial core. We characterize the operational boundaries of the proposed architectures by varying four key parameters: Data Volume, Channel Quality, Model Complexity, and Environmental Stability.
\rev{In this section, we numerically evaluate the analytical feasibility framework derived in Section~V. The objective is to characterize the energy and latency boundaries of three representative O-RAN AI placements under common workload and communication assumptions, rather than to benchmark a particular online scheduling algorithm. We consider Ground-Centric execution (S1) as the conventional centralized-processing reference, Ground--LEO Split-RIC execution (S2) as the local-inference alternative, and GEO--LEO execution (S3) as the ISL-assisted model-refresh alternative. Numerical sensitivity analyses are performed over telemetry volume, feeder-link rate, inference complexity, model longevity, orbital waiting time, and model-update requirements.}

\subsection{Simulation Setup}
\label{sim_setup}
The simulation parameters, summarized in Table \ref{tab:SimParams}, represent reference operating conditions with a Ka-band LEO--Ground feeder link and an optical GEO--LEO ISL for the multi-layer scenario. The hardware profiles match the definitions in Section \ref{sec:SystemModel}.

\rev{The evaluation is performed at the level of a representative communication and computation path, rather than through a constellation-wide orbital simulation. The modeled entities comprise one LEO inference node, one Ground gateway connected to a terrestrial computing facility, and, for S3, one GEO training node. S1 and S2 use the LEO--Ground feeder link, whereas S3 uses the LEO--GEO ISL. The resource and link parameters are specified in Tables~\ref{tab:NodeResources} and \ref{tab:SimParams}, respectively.}

\begin{table}[tbp]
\caption{Simulation Parameters}
\label{tab:SimParams}
\centering
\resizebox{\columnwidth}{!}{%
\begin{tabular}{@{}llc@{}}
\toprule
\textbf{Parameter} & \textbf{Symbol} & \textbf{Value} \\
\midrule
\multicolumn{3}{c}{\textit{Orbital \& Link Parameters}} \\
Orbit Altitude & $h_{LEO}$ & \qty{600}{\km} \\
%Feeder Link Frequency & $f_c$ & \qty{28}{\giga\hertz} (Ka-band) \\
\rev{Feeder UL/DL Rate} & \rev{$R_{\mathrm{ul}}^{\mathrm{feed}}$, $R_{\mathrm{dl}}^{\mathrm{feed}}$} & \rev{\qty{500}{\mega\bit\per\second}}\\
% Feeder Link Bandwidth & $B_{feed}$ & \qty{400}{\mega\hertz} \\
\rev{Feeder Link RTT} & \rev{$RTT_{\mathrm{feed}}$} & \rev{\qty{20}{\milli\second}}\\
Feeder Tx Power & $P_{tx}^{feed}$ & \qty{15}{\watt} \\
Feeder Rx Power & $P_{rx}^{feed}$ & \qty{5}{\watt} \\
ISL Tx/Rx Power (Optical) & $P_{tx}^{ISL},P_{rx}^{ISL}$ & \qty{2}{\watt} \\
ISL Data Rate & $R_{ISL}$ & \qty{10}{\giga\bit\per\second} \\
Orbital Waiting Time & $T_{wait}$ & \qtyrange[range-units = single, range-phrase = --]{0}{60}{\minute} \\
\rev{Ground Effective Throughput} & \rev{$F_{\mathrm{GND}}$} & \rev{1 PFLOP/s} \\
\rev{LEO Power-Feasible Throughput} & \rev{$F_{\mathrm{LEO}}^{\mathrm{eff}}$} & \rev{1 TFLOP/s} \\
\rev{GEO Power-Feasible Throughput} & \rev{$F_{\mathrm{GEO}}^{\mathrm{eff}}$} & \rev{10 TFLOP/s} \\
\midrule
\multicolumn{3}{c}{\textit{AI Workload Profile}} \\
Input Data Size (Variable) & $\delta^{in}$ & \qtyrange[range-units = single, range-phrase = --]{10}{50000}{\kilo\byte} \\
\rev{Output Data Size} & \rev{$\delta^{\mathrm{out}}$} & \rev{\qty{1}{\kilo\byte}} \\
Inference \rev{Workload} (Variable) & $\omega^{inf}$ & \qtyrange[range-units = single, range-phrase = --]{0.1}{500}{\giga\text{FLOP}\per\text{inference}} \\
Training Dataset Size & $\Delta^{train}$ & \qty{10}{\giga\bit} \\
\rev{Training Workload} & \rev{$\Omega^{\mathrm{train}}$} & \rev{3 PFLOP} \\
Model Weight Size & $\sigma$ & \qty{50}{\mega\bit} \\
Model Longevity & $N_{inf}$ & $10^5$ inference events \\ 

\bottomrule
\end{tabular}%
}
\end{table}

\rev{No specific commercial constellation, orbital phasing, or geographic distribution of ground stations is reproduced. Instead, ground-contact intermittency is represented explicitly by the waiting-time parameter $T_{\mathrm{wait}}$, swept from $0$ to $60$~min. This permits the evaluation of different contact conditions without attributing a particular waiting time to an unspecified constellation. The baseline ISL-assisted scenario assumes a link available at the refresh trigger; additional acquisition and availability constraints are discussed separately. All comparisons use identical training and inference workloads, so that the observed differences arise from the considered functional placement and communication path.}

\rev{For the reference parameters in Table~\ref{tab:SimParams}, the fixed contribution to the S2 model-refresh latency is $23.12$~s. Accordingly, the baseline distinct-pass assumption, in which training-data upload and model return require separate ground-contact opportunities, yields $T_{\mathrm{refresh},S2}^{\mathrm{baseline}}=2T_{\mathrm{wait}}+23.12$~s. Under the favorable same-pass assumption, the updated model is returned without an additional orbital waiting interval, yielding $T_{\mathrm{refresh},S2}^{\mathrm{same-pass}}=T_{\mathrm{wait}}+23.12$~s. Both cases are considered in the numerical evaluation to assess the impact of ground-contact availability. The inference output is set to $\delta^{out}=1$~kB, representing a compact control result.}

%\rev{The S2 curves report both the baseline store-and-forward case, $T_{\mathrm{refresh},S2}^{\mathrm{baseline}}=2T_{\mathrm{wait}}+23.12$~s, and the favorable same-pass return case, $T_{\mathrm{refresh},S2}^{\mathrm{same\mbox{-}pass}}=T_{\mathrm{wait}}+23.12$~s, thereby showing the sensitivity of the model-refresh latency to the assumed ground-contact pattern.}

\rev{The numerical comparison is intentionally architecture-level rather than algorithm-level. Published approaches such as PHOENIX \cite{Liu2024Phoenix} optimize sunlight-aware task scheduling and battery consumption, whereas LEOEdge~\cite{10694785} addresses distributed AI inference placement and delivery latency. Their performance metrics and resource models differ from the lifecycle energy and model-refresh metrics considered here. Consequently, we do not claim algorithmic superiority over these approaches; a direct comparison would require implementing the alternative schedulers within a common constellation, workload, and energy model.} %\rev{The purpose of the comparison is therefore to identify architecture-dependent feasibility boundaries under common workload assumptions, rather than to establish performance gains over independently optimized scheduling algorithms.}

\subsection{Energy Sensitivity Analysis}

\subsubsection{Communication Constraints}
\begin{figure}[tbp]
    \centering
    \includegraphics[width=0.9\columnwidth]{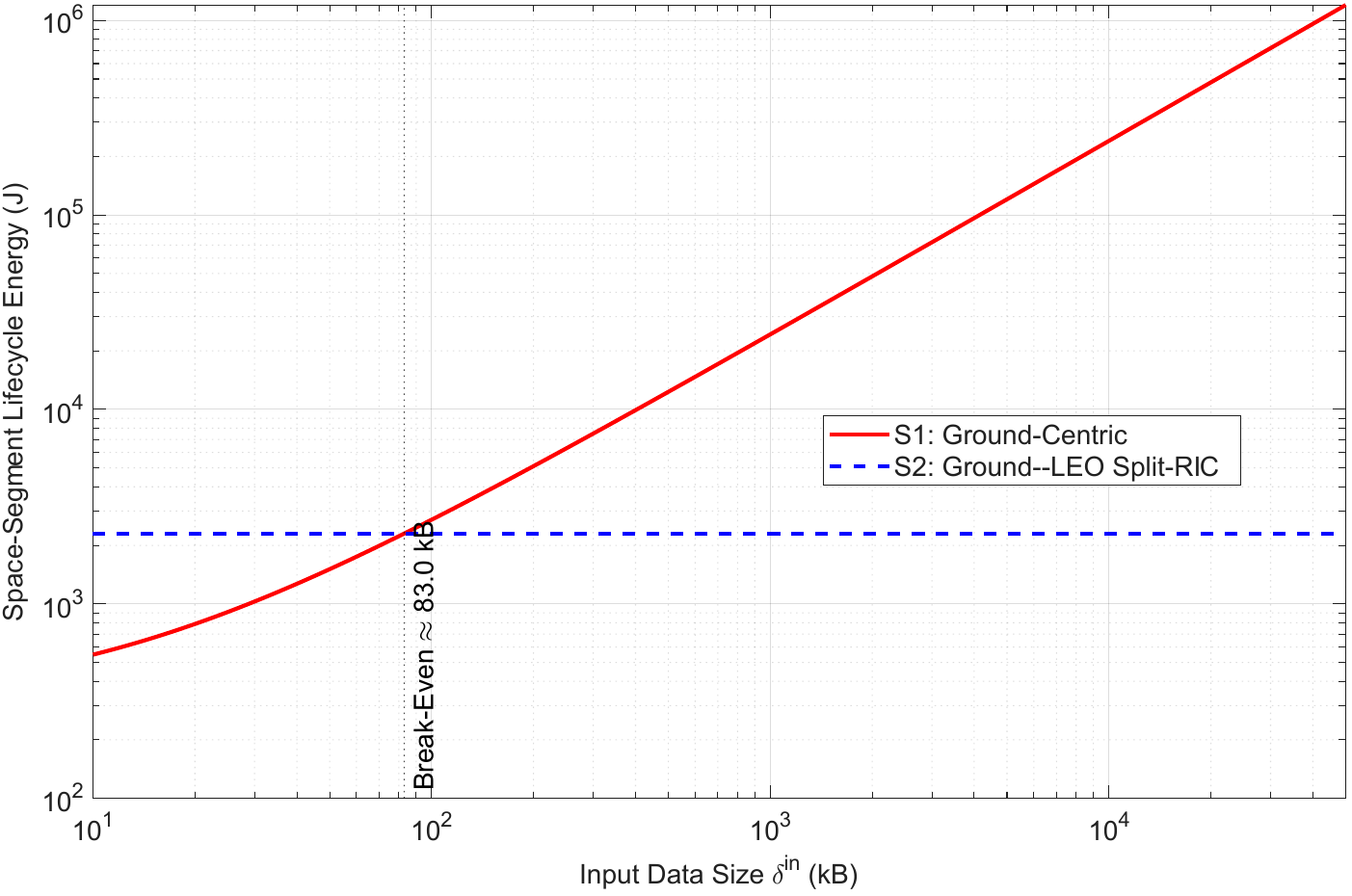}
    \caption{\rev{Space-segment lifecycle energy versus input telemetry size for S1 and S2, with $\omega^{\mathrm{inf}}=1$~GFLOP/inference. %The exact energy break-even occurs at $\delta^{\mathrm{in}}\simeq83$~kB.
    }}
    \label{fig:energy_data}
\end{figure}

We first \rev{investigate} how the \rev{telemetry volume and} feeder-link \rev{conditions affect} the \rev{space-segment lifecycle energy of S1 and S2}. Fig.~\ref{fig:energy_data} compares the \rev{two architectures} as a function of the \rev{input} telemetry size \rev{$\delta^{\mathrm{in}}$}, for a fixed \rev{inference workload} $\omega^{inf} = \qty{1}{\giga\text{FLOP}\per\text{inference}}$. \rev{The Ground-Centric architecture is preferable for small telemetry messages, since offloading a limited amount of data requires less space-segment energy than executing the inference workload on-board. As $\delta^{\mathrm{in}}$ increases, however, the repeated feeder-link transmission required by S1 becomes dominant, whereas the S2 energy remains essentially independent of the inference-input size because the telemetry is processed locally on the LEO node.}

\rev{Using the exact lifecycle-energy expressions in Section~\ref{sec:Analysis}, the S1--S2 energy break-even occurs at $\delta^{\mathrm{in}}\simeq83.0$~kB for the considered 1-GFLOP/inference workload. Above this threshold, S2 is energetically preferable. For example, at $\delta^{\mathrm{in}}=5$~MB, the lifecycle energies are approximately $120.3$~kJ for S1 and $2.30$~kJ for S2, corresponding to an energy reduction of about $98\%$. The crossover should be interpreted as an energy-preference boundary rather than as a computational-feasibility limit.}

\begin{figure}
    \centering
    \includegraphics[width=0.9\columnwidth]{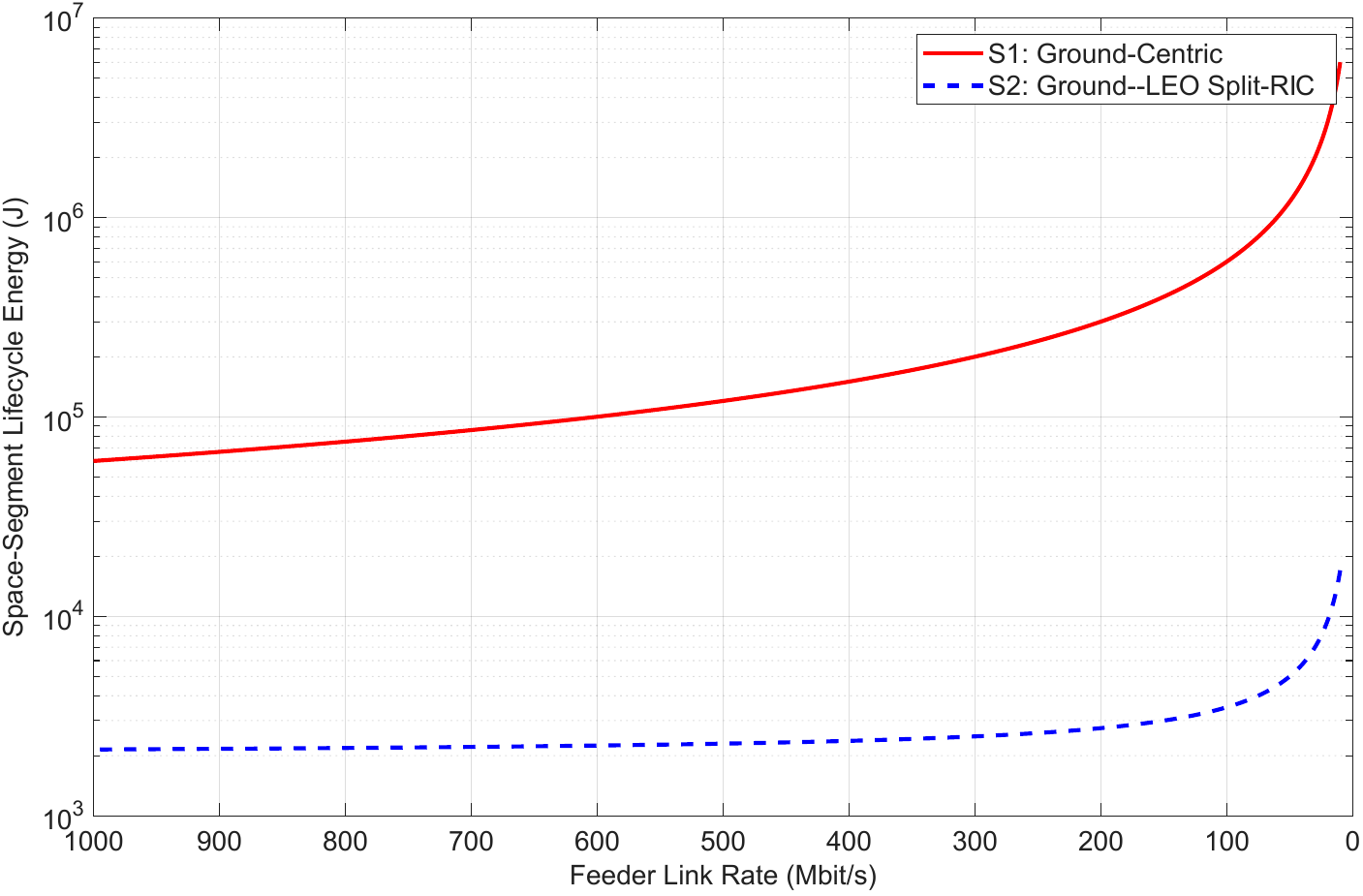}
    \caption{\rev{Space-segment lifecycle energy versus feeder-link rate for S1 and S2, with $\delta^{\mathrm{in}}=5$~MB and $\omega^{\mathrm{inf}}=1$~GFLOP/inference. %S2 is substantially less sensitive to feeder-link degradation because inference is executed locally on the LEO node.
    }}
    \label{fig:energy_rate}
\end{figure}

\rev{Fig.~\ref{fig:energy_rate} evaluates the same architectures for a fixed $\delta^{\mathrm{in}}=5$~MB and $\omega^{\mathrm{inf}}=1$~GFLOP/inference while varying the feeder rate. Both architectures become more energy demanding as the link rate decreases because the periodic training-data transfer remains necessary. S1 is considerably more sensitive, however, since every inference event additionally requires transmission of the high-volume input telemetry. At $1$~Gbit/s, S1 and S2 consume approximately $60.15$~kJ and $2.15$~kJ, respectively, whereas at $50$~Mbit/s the corresponding values increase to approximately $1.20$~MJ and $5.01$~kJ. The analytical S1--S2 rate crossover is approximately $30$~Gbit/s, well above the evaluated $10$~Mbit/s--$1$~Gbit/s range; consequently, S2 remains the lower-energy solution throughout the considered feeder-link conditions.}

%Fig. \ref{fig:EnergyVsRate} evaluates a critical practical scenario: rain attenuation or low-elevation angles reducing the Feeder Link rate ($R_{ul}$). We assume a \qty{5}{\mega\byte} Spectrogram xApp. The energy cost of S1 is inversely proportional to the data rate ($E \propto 1/R_{ul}$). As the channel degrades (e.g., dropping from \qty{1}{\giga\bit\per\second} to \qty{50}{\mega\bit\per\second}), the transmission time extends, causing energy consumption to explode. In contrast, the Split-RIC architecture (S2) is less immune to channel degradation, confirming it as the robust choice for high-latitude or weather-affected regions.

\subsubsection{Computational Constraints} 
\begin{figure}
    \centering
    \includegraphics[width=0.9\columnwidth]{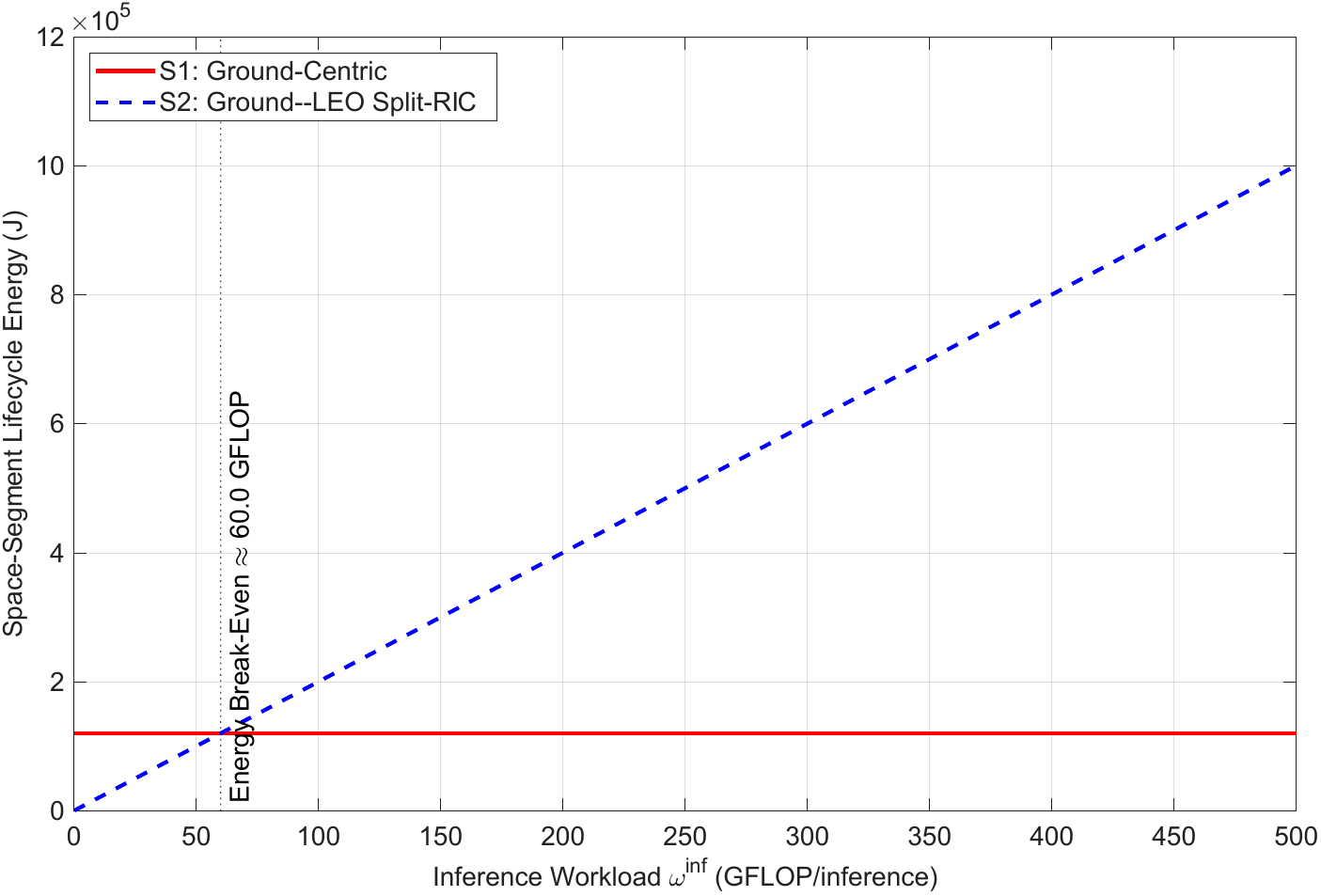}
    \caption{\rev{Space-segment lifecycle energy versus inference workload for S1 and S2, with $\delta^{\mathrm{in}}=5$~MB. %The exact energy-preference crossover occurs at $\omega^{\mathrm{inf}}\simeq60$~GFLOP/inference.
    }}
    \label{fig:energy_complexity}
\end{figure}

\rev{We next investigate how the inference workload affects the space-segment lifecycle-energy preference between S1 and S2. Fig.~\ref{fig:energy_complexity} considers a fixed telemetry input of $\delta^{\mathrm{in}}=5$~MB and varies the inference workload $\omega^{\mathrm{inf}}$ over the range $0.1$--$500$~GFLOP/inference. Since inference is performed on the Ground Segment in S1, its space-segment lifecycle energy is independent of $\omega^{\mathrm{inf}}$ and remains approximately $120.3$~kJ. In contrast, the S2 energy increases linearly with the on-board inference workload through the term $N_{\mathrm{inf}}\omega^{\mathrm{inf}}\epsilon_{\mathrm{LEO}}$.}

%Next, we analyze the limits of the on-board hardware. Fig. \ref{fig:EnergyVsComplexity} investigates the upper bound of on-board AI. We fix the data size at $\delta^{in} = \qty{5}{\mega\byte}$ and vary the computational complexity of the neural network. The energy cost of S1 is flat. Since processing occurs on the ground (where power is considered infinite/free relative to the battery), increasing model depth does not impact the satellite's energy budget. The energy cost of S2 rises linearly with FLOPs. At approximately $\omega^{inf} \approx \qty{250}{\giga\text{FLOPS}}$, the on-board computation cost exceeds the transmission cost. This defines the \textit{Physical Feasibility Boundary}: if a specific xApp requires a massive Transformer model exceeding this complexity, it must be offloaded to the ground (S1) or a powerful GEO hub (S3), regardless of the data transmission penalty.

\rev{The exact S1--S2 energy break-even occurs at $\omega^{\mathrm{inf}}\simeq60.0$~GFLOP/inference for the considered $5$~MB telemetry input. Below this threshold, local LEO inference is energetically preferable, whereas above it the communication cost of S1 becomes lower than the accumulated on-board inference energy of S2. Importantly, this crossover is an energy-preference boundary and should not be interpreted as a hardware or computational-feasibility ceiling. Computational feasibility is instead determined by the control-loop condition $\omega^{\mathrm{inf}}/F_{\mathrm{LEO}}^{\mathrm{eff}} \leq\tau^{\max}$ derived in Section~\ref{sec:Analysis}.}

\begin{figure}
    \centering
    \includegraphics[width=0.9\columnwidth]{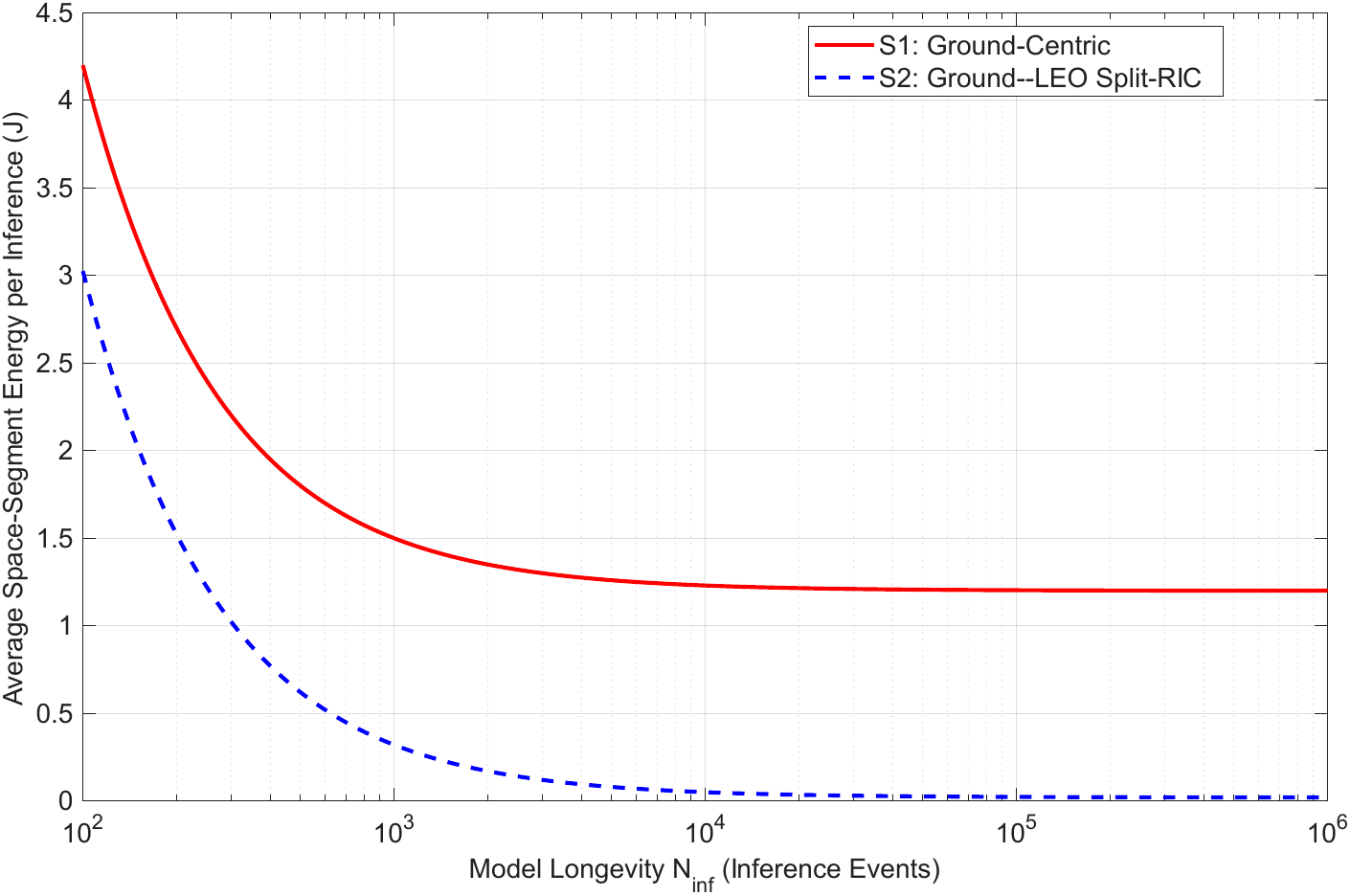}
    \caption{\rev{Average space-segment energy per inference versus model longevity for S1 and S2, with $\delta^{\mathrm{in}}=5$~MB and $\omega^{\mathrm{inf}}=1$~GFLOP/inference. %The update overhead is progressively amortized as $N_{\mathrm{inf}}$ increases, while S2 remains energetically preferable throughout the considered range.
    }}
    \label{fig:energy_longevity}
\end{figure}

\rev{Fig.~\ref{fig:energy_longevity} evaluates the effect of model longevity $N_{\mathrm{inf}}$ by reporting the average space-segment energy per inference for the same $\delta^{\mathrm{in}}=5$~MB and $\omega^{\mathrm{inf}}=1$~GFLOP/inference operating point. Shorter model lifetimes increase the amortized contribution of the one-time training-data and model-distribution overheads. However, under the revised lifecycle-energy accounting, S2 remains energetically preferable to S1 throughout the considered $10^2$--$10^6$ inference-event range. The analytical crossover would occur at only approximately $0.42$ inference events and is therefore outside the physically meaningful operating region.}

%Fig. \ref{fig:EnergyVsLongevity} analyzes the \textit{Amortized Energy per Inference}. If the environment is unstable and the model requires retraining every few minutes (Low $N_{inf}$), the overhead of receiving model updates (via A1 interface) makes S2 inefficient. S2 is only feasible for stable environments where a model remains valid for $> 1000$ inference events. 

\rev{At $N_{\mathrm{inf}}=100$, the average space-segment energy is approximately $4.20$~J/inference for S1 and $3.03$~J/inference for S2. As $N_{\mathrm{inf}}$ increases and the one-time update overhead is progressively amortized, the curves approach their per-inference asymptotes of approximately $1.200$~J for S1 and $0.020$~J for S2. Thus, model longevity affects the magnitude of the energy advantage, but does not change the preferred architecture over the evaluated range for this workload.}

\subsection{\rev{Model-Refresh Latency Analysis}}

\rev{We next evaluate the model-refresh latency introduced in
Section~V, i.e., the time between an update trigger and the
availability of the refreshed model at the node executing
inference. This quantity is distinct from the per-inference control
latency and therefore does not accumulate over the
$N_{\mathrm{inf}}$ inference events.}

\begin{figure}
    \centering
    \includegraphics[width=0.9\columnwidth]{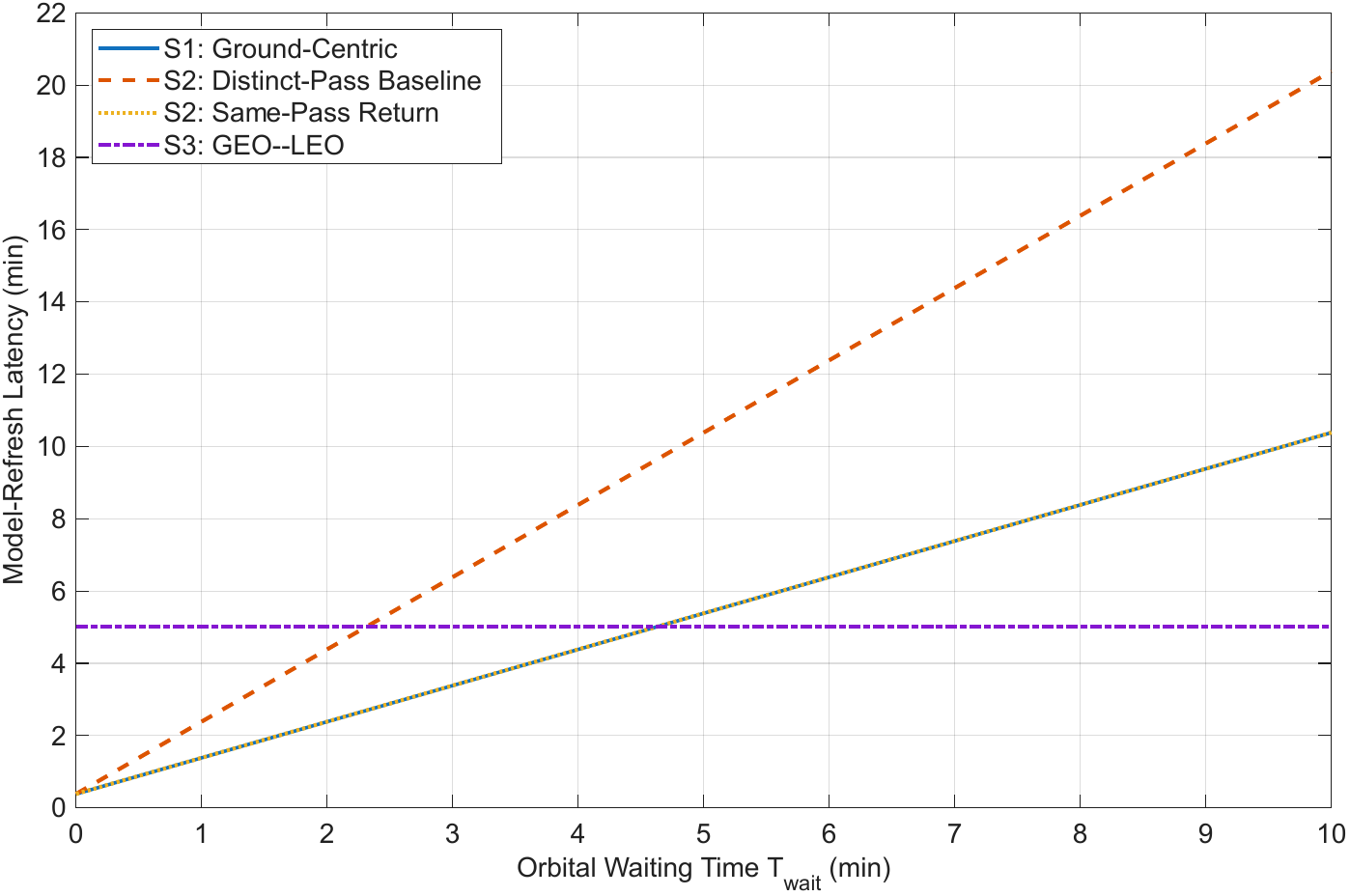}
    \caption{\rev{Model-refresh latency versus orbital waiting time. %The numerical analysis spans the full $0$--$60$~min range of Table~\ref{tab:SimParams}, while the figure focuses on the $0$--$10$~min interval containing all pairwise latency crossovers. S2 is shown for both the baseline distinct-pass and the favorable same-pass model-return assumptions.
    }}
    \label{fig:latency_wait}
\end{figure}

\rev{Fig.~\ref{fig:latency_wait} reports the model-refresh latency
as a function of the orbital waiting time $T_{\mathrm{wait}}$.
The numerical analysis considers the full $0$--$60$~min range in
Table~\ref{tab:SimParams}, while the figure focuses on the
$0$--$10$~min interval containing all pairwise latency
crossovers. For the adopted parameters, the refresh latencies are $
T_{\mathrm{refresh},S1}
= T_{\mathrm{wait}} + 23.01~\mathrm{s}$, $T_{\mathrm{refresh},S2}^{\mathrm{baseline}}
= 2T_{\mathrm{wait}} + 23.12~\mathrm{s}$, $T_{\mathrm{refresh},S2}^{\mathrm{same\mbox{-}pass}}= T_{\mathrm{wait}} + 23.12~\mathrm{s}$, and $T_{\mathrm{refresh},S3}= 301.24~\mathrm{s}$.
}

%
%
%
% \subsection{Latency Feasibility Analysis}
% While S2 offers energy savings, it introduces latency challenges. We analyze the Time-Criticality of the architectures.
%
% Fig. \ref{fig:LatencyVsWait} plots the Total Lifecycle Latency ($\mathcal{T}$) against the average orbital waiting time ($T_{wait}$), which depends on the density of the ground station network. The latency of the Ground-LEO split is dominated by the store-and-forward delay ($2 \times T_{wait}$). In sparse deployments where $T_{wait} \approx \qty{45}{\minute}$, the model update cycle is too slow for dynamic threats (e.g., jamming adaptation). The Multi-Layer GEO-LEO architecture maintains a constant, low latency ($\mathcal{T}_{S3} \approx \qty{2}{\second}$) regardless of the LEO satellite's position. As a result, S3 is the mandatory architecture for \textit{Time-Critical Adaptation}. S2 is only feasible for \textit{Stationary Optimization} tasks (e.g., daily traffic pattern learning) where update delays are acceptable.
%
%
\rev{At small orbital waiting times, the Ground-based training
options provide the shortest model-refresh latency despite the
additional feeder-link transfer. In particular, S1 is always
slightly faster than the favorable same-pass realization of S2
because the latter additionally requires redistribution of the
updated model to the LEO node. Their fixed difference is only
approximately $0.11$~s, which explains the near-overlap of the two
curves in Fig.~\ref{fig:latency_wait}.}

\rev{The S3 refresh latency is independent of ground-station
visibility in the considered reference geometry, but it includes
the GEO training time. With
$\Omega^{\mathrm{train}}=3$~PFLOP and
$F_{\mathrm{GEO}}^{\mathrm{eff}}=10$~TFLOP/s, this contribution is
$300$~s and dominates the resulting
$T_{\mathrm{refresh},S3}\simeq5.02$~min. Consequently, S3 becomes
faster than the S2 distinct-pass baseline only when
$T_{\mathrm{wait}}\gtrsim2.32$~min. For the favorable same-pass
case, the crossover moves to approximately $4.64$~min, while the
corresponding S1--S3 crossover is also approximately
$4.64$~min.}

\rev{These results highlight that the value of S3 is not a
universally lower refresh latency, but rather its independence from
ground-contact waiting. Hence, GEO-assisted training becomes
latency-attractive when orbital access delays are sufficiently
large, whereas Ground-based training remains preferable when a
gateway is promptly available. This comparison concerns the model
refresh process only: the near-real-time control latency of S2 and
S3 is identical in the adopted architecture because inference is
executed locally on the LEO node in both cases.}

\subsection{Multi-Dimensional Feasibility Maps}
\label{maps}
\begin{figure}
    \centering
    \includegraphics[width=0.9\columnwidth]{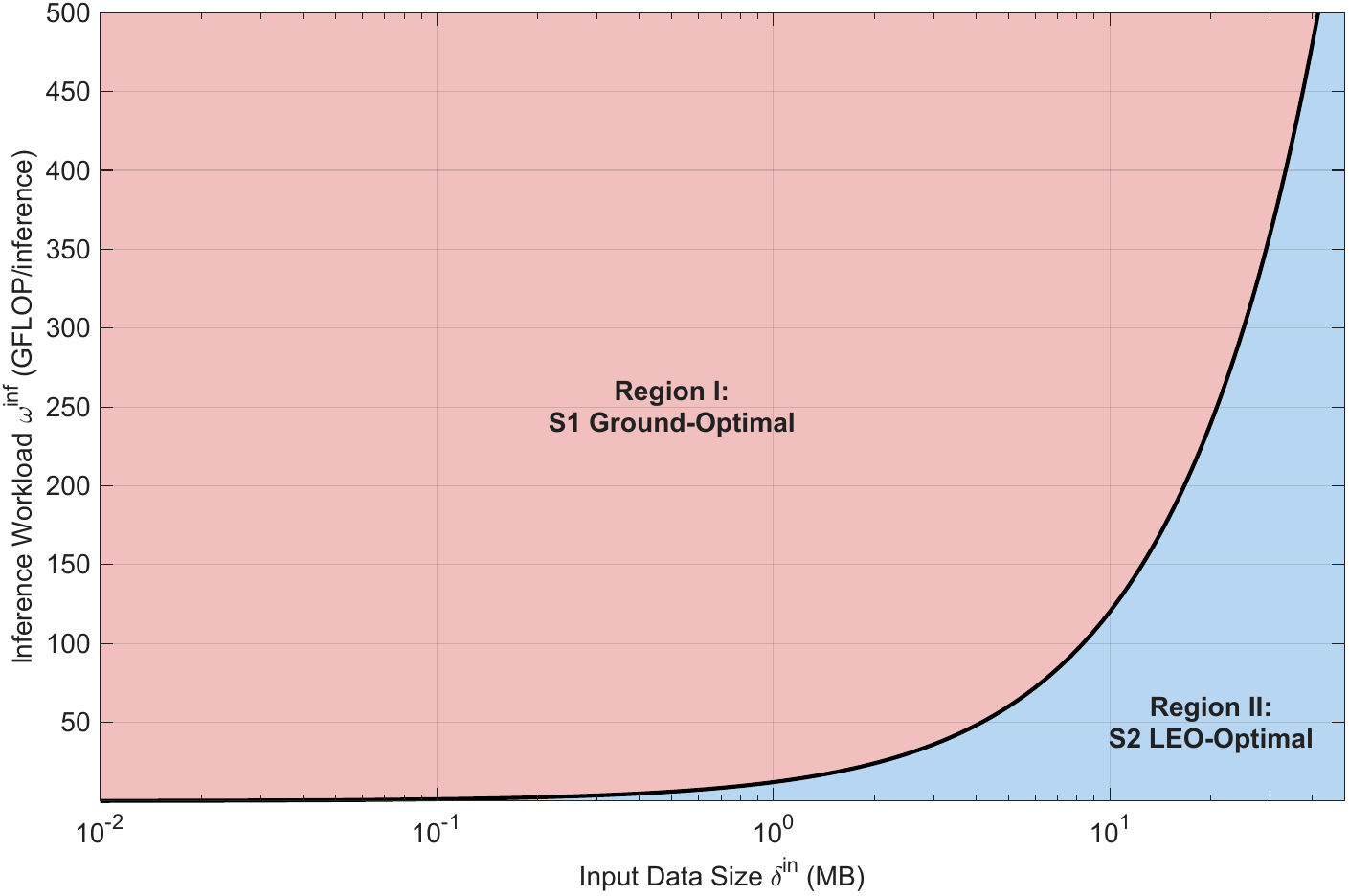}
    \caption{\rev{Space-segment energy-feasibility map as a function of input telemetry size $\delta^{\mathrm{in}}$ and inference workload $\omega^{\mathrm{inf}}$. %Region~I identifies operating points where S1 is energy-optimal, whereas Region~II identifies those where S2 is energy-optimal under the baseline lifecycle parameters.
    }}
    \label{fig:energy_map}
\end{figure}

\rev{We finally combine the previous one-dimensional sensitivity analyses into feasibility maps that jointly expose the operating conditions governing the architecture choice. Fig.~\ref{fig:energy_map} shows the space-segment energy preference between S1 and S2 as a function of the input telemetry size $\delta^{\mathrm{in}}$ and the inference workload $\omega^{\mathrm{inf}}$. The map is obtained from the exact lifecycle-energy expressions in Section~V, including the one-time training-data transfer, model distribution, and control-output terms.}

%Finally, we synthesize these trade-offs into two unified decision maps to guide network orchestration. The Energy Map (Fig. \ref{fig:MapEnergy}) is a contour plot delineating the boundary between Ground-Optimal (Region I) and LEO-Optimal (Region~II). Network operators can map xApps to this plane: Traffic Prediction (Low Data, High Compute) falls in Region~I, while Beam Management (High Data, Low Compute) falls in Region II. The Latency Map plot (Fig. \ref{fig:MapLatency}) highlights the necessity of Scenario 3. As the required update frequency increases (Y-axis) and ground station density decreases (X-axis, higher wait time), the system enters the Dark Zone where S2 violates latency deadlines. In this region, the Multi-Layer GEO-LEO architecture (S3) is not just an option, but a physical requirement.

\rev{Region~I identifies operating points for which S1 is energy-optimal. This region is favored by relatively small telemetry volumes and/or high inference workloads, for which transmitting the input to the Ground Segment costs less space-segment energy than executing inference repeatedly on-board. Region~II identifies the complementary regime in which S2 is energy-optimal: as the telemetry volume increases, avoiding its repeated transmission over the feeder link compensates for the energy required by local LEO inference. The boundary therefore represents an energy-preference crossover rather than a computational-feasibility limit.}

\rev{S3 does not form an energy-optimal region under the baseline parameters considered here. S2 and S3 incur the same LEO inference energy, whereas S3 additionally performs the $\Omega^{\mathrm{train}}=3$~PFLOP training workload at the GEO node and exchanges the training dataset and updated model over the ISL. With the parameters of Table~\ref{tab:SimParams}, this results in an approximately $300$~kJ GEO training contribution, and $J_{S3}^{\mathrm{space}}-J_{S2}^{\mathrm{space}}\simeq299.7$~kJ, independently of $\delta^{\mathrm{in}}$ and $\omega^{\mathrm{inf}}$. Hence, the role of S3 in the considered baseline is not to minimize energy, but to remove the dependence of model refresh on ground-station visibility, as investigated next.}

\begin{figure}
    \centering
    \includegraphics[width=0.9\columnwidth]{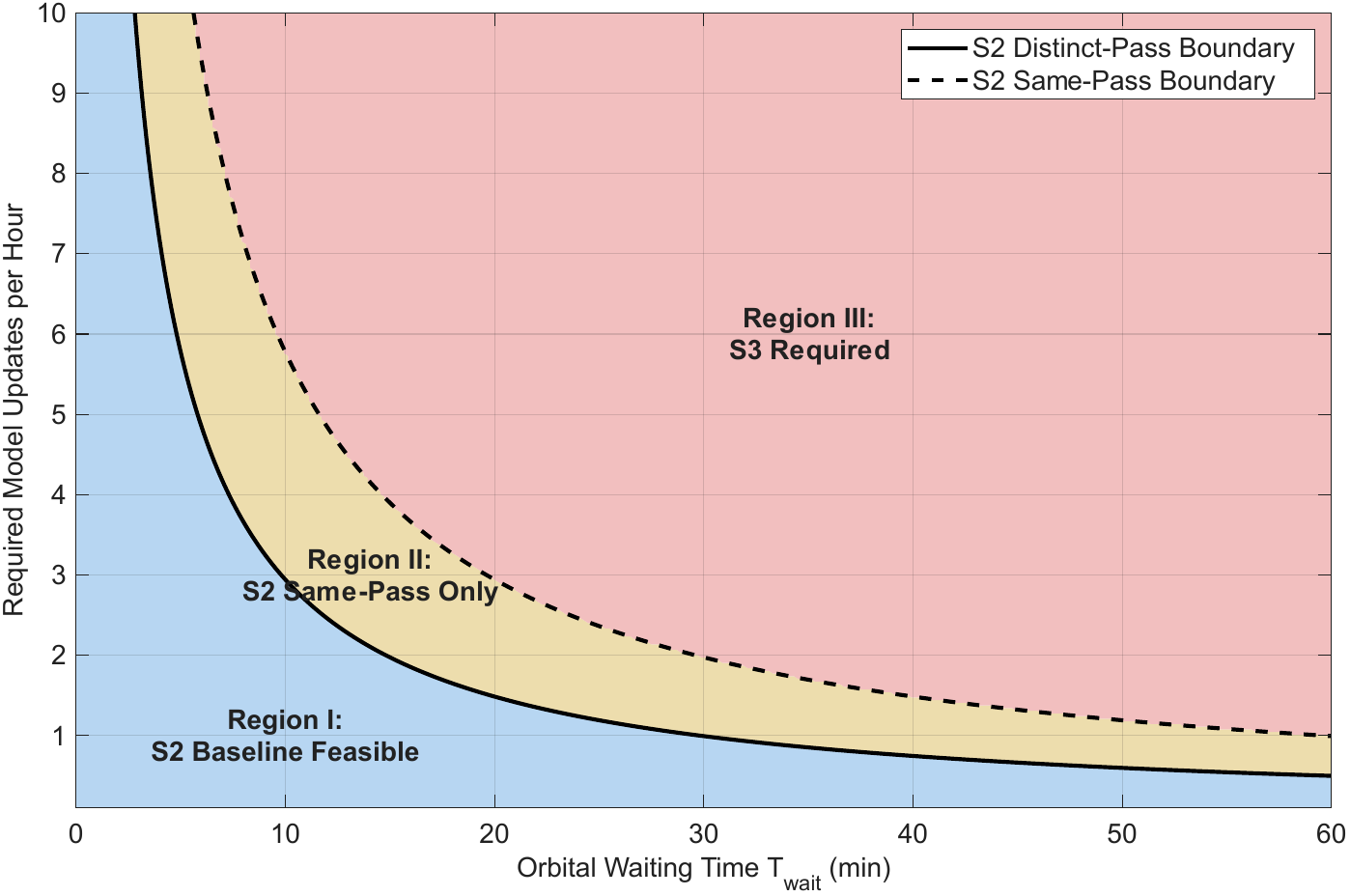}
    \caption{\rev{Model-refresh feasibility map versus orbital waiting time and required model-update frequency. %Region~I denotes conditions under which the S2 distinct-pass baseline satisfies the refresh deadline; Region~II denotes conditions where only the favorable S2 same-pass realization remains feasible; Region~III denotes conditions where S3 is required, among the considered local-inference architectures, to satisfy the deadline.
    }}
    \label{fig:latency_map}
\end{figure}

\rev{Fig.~\ref{fig:latency_map} complements the energy map by showing model-refresh feasibility as a joint function of the orbital waiting time $T_{\mathrm{wait}}$ and the required model-update frequency. A requirement of $f_{\mathrm{upd}}$ updates/hour is translated into the refresh-latency constraint
\begin{equation}
T_{\mathrm{refresh}}
\leq
\tau_{\mathrm{upd}}
=
\frac{3600}{f_{\mathrm{upd}}},
\end{equation}
so that increasing the required update frequency corresponds to a
progressively tighter refresh deadline.}

\rev{Region~I identifies the operating points for which the baseline S2 distinct-pass realization satisfies the required refresh deadline. In this region, the complete Ground--LEO update cycle, including separate waiting intervals for training-data upload and model return, remains sufficiently fast. Region~II corresponds to the intermediate regime in which the distinct-pass realization no longer satisfies the deadline, while S2 remains feasible if the updated model can be returned during the same ground-station contact. This region therefore explicitly quantifies the benefit of a favorable contact schedule rather than treating orbital waiting time as a single deterministic penalty.}

\rev{Region~III occurs when neither S2 realization can satisfy the model-refresh requirement. Under these conditions, S3 removes the ground-contact waiting term by performing training through the GEO-assisted path and is therefore required, among the considered local-inference architectures, to meet the refresh deadline. For the baseline parameters, $T_{\mathrm{refresh},S3}=301.24$~s, corresponding to a maximum supported update frequency of approximately $11.95$~updates/hour. Since Fig.~\ref{fig:latency_map} considers requirements up to $10$~updates/hour, S3 satisfies the refresh deadline throughout the displayed range.}

\rev{For example, with a requirement of $10$~updates/hour, corresponding to a $6$-min refresh deadline, the distinct-pass S2 realization remains feasible only for $T_{\mathrm{wait}}\lesssim2.81$~min, while the same-pass realization extends this limit to approximately $5.61$~min. For a slower requirement of $1$~update/hour, the corresponding waiting-time limits increase to approximately $29.81$~min and $59.61$~min, respectively. Hence, the architectural transition is jointly determined by the required adaptation rate and the availability of ground contacts.}

\rev{The two maps lead to the following architecture-selection guidelines. First, S1 is favored from a space-segment energy perspective when telemetry volumes are small and/or inference workloads are sufficiently high. Second, S2 is favored when repeated telemetry transmission dominates the lifecycle energy and the required model-refresh deadline can still be satisfied through available ground contacts. Third, S3 becomes relevant when ground-contact intermittency prevents S2 from satisfying the required model-refresh deadline, accepting the additional GEO training-energy cost in exchange for contact-independent model refresh.}

\rev{As an illustrative joint-feasibility comparison, consider $\delta^{\mathrm{in}}=5$~MB, $\omega^{\mathrm{inf}}=1$~GFLOP/inference, $T_{\mathrm{wait}}=5$~min, a $10$-ms control deadline, and a required model-refresh interval of $360$~s. Under the baseline parameters, S1, S2, and S3 respectively consume approximately $120.31$, $2.30$, and $302.00$~kJ of space-segment lifecycle energy. However, S1 violates the control deadline ($T_{\mathrm{ctrl},S1}\approx100$~ms), whereas S2 and S3 satisfy it ($T_{\mathrm{ctrl}}\approx1$~ms). For S2, the distinct-pass refresh delay is approximately $623$~s and therefore violates the update deadline, while its favorable same-pass realization requires approximately $323$~s and remains feasible. S3 provides approximately $301$~s under the baseline ISL assumption. Thus, S3 is the only feasible deployment among these alternatives when S2 requires distinct contacts, whereas S2 is energetically preferable if same-pass model delivery is possible. This illustrates the need to evaluate energy preference jointly with control and model-refresh feasibility.}

% We can synthesize these results into a unified a set of guidance rules.
% \begin{enumerate}
% \item \textbf{Region I (Ground-Optimal):} \textit{Low Data Volume.} Tasks involving simple reporting should use the legacy O1 interface (S1).
% \item \textbf{Region II (LEO-Optimal):} \textit{High Data Volume + Moderate Complexity.} Tasks involving raw signal processing (e.g., Beam Management) should be executed on the LEO Edge (S2) to maximize energy efficiency.
% \item \textbf{Region III (GEO-Optimal):} \textit{High Agility Requirement.} Tasks requiring immediate model retraining must utilize the GEO ISL backbone (S3), accepting a higher energy cost for the sake of speed.
% \end{enumerate}

\subsection{\rev{Parameterization and Scope of Validity}}\label{sec:Validation}

\rev{The parameters in Table~\ref{tab:SimParams} define representative operating conditions rather than a specific satellite platform or constellation. Spaceborne processing throughputs are constrained by~\eqref{eq:effective_throughput}, yielding $F_{\mathrm{LEO}}^{\mathrm{eff}}=1$~TFLOP/s and $F_{\mathrm{GEO}}^{\mathrm{eff}}=10$~TFLOP/s, despite nominal peak values of 10 and 200~TFLOP/s. The Ground reference throughput is $F_{\mathrm{GND}}=1$~PFLOP/s, with terrestrial computational resources assumed non-binding in the adopted space-segment feasibility model.} \rev{The reference feeder link uses $R_{\mathrm{ul}}^{\mathrm{feed}}=R_{\mathrm{dl}}^{\mathrm{feed}}=500$~Mbit/s and $\mathrm{RTT}_{\mathrm{feed}}=20$~ms, whereas the GEO--LEO ISL uses $R_{\mathrm{ISL}}=10$~Gbit/s and $\mathrm{RTT}_{\mathrm{ISL}}\simeq235$~ms for a favorable reference geometry. These parameters characterize an available link during the model-refresh operation, not permanent GEO--LEO connectivity.} \rev{The baseline workload assumes $\sigma=50$~Mbit, $\Omega^{\mathrm{train}}=3$~PFLOP, and $\delta^{\mathrm{out}}=1$~kB. The training workload corresponds to approximately 3~s on the Ground reference platform and 300~s on the GEO node. The inference workload is varied over $0.1$--$500$~GFLOP/inference to characterize energy-preference boundaries rather than specific xApp implementations. The sensitivity to alternative workloads, connectivity assumptions, and hardware performance is discussed next.}

\subsection{\rev{Robustness to Modeling Assumptions}}

\rev{The feasibility boundaries are parameter-dependent and can incorporate alternative modeling assumptions. If telemetry compression reduces the transmitted volume to $\delta_{\mathrm{tx}}^{\mathrm{in}}=\kappa_\delta\delta_{\mathrm{raw}}^{\mathrm{in}}$, with $\kappa_\delta\in(0,1]$, the S1--S2 energy break-even, expressed in terms of raw input volume, increases proportionally to $1/\kappa_\delta$. Thus, compression extends the region in which Ground-based inference remains energetically competitive.}

\rev{Multiple training or model-exchange rounds can be represented through aggregate quantities $\Delta_{\mathrm{eff}}^{\mathrm{train}}$, $\sigma_{\mathrm{eff}}$, and $\Omega_{\mathrm{eff}}^{\mathrm{train}}$, replacing the corresponding single-cycle parameters in Section~\ref{sec:Analysis}. Larger model-transfer requirements penalize architectures requiring model dissemination, whereas increased training complexity particularly affects S3 through GEO computation energy and delay.}

\rev{Intermittent GEO--LEO connectivity introduces an additional ISL waiting time $T_{\mathrm{wait}}^{\mathrm{ISL}}$ into $T_{\mathrm{refresh},S3}$. For the S2 distinct-pass baseline, the corresponding ground-contact waiting-time crossover increases by $T_{\mathrm{wait}}^{\mathrm{ISL}}/2$. Ground-station availability is represented by the $0$--$60$~min sweep of $T_{\mathrm{wait}}$ in Figs.~\ref{fig:latency_wait} and~\ref{fig:latency_map}; mapping an actual gateway distribution to this parameter requires orbital analysis beyond the present model.}

\rev{Uncertainty in sustained computational throughput can be expressed as $\alpha_n F_n^{\mathrm{eff}}$, with $\alpha_n\in(0,1]$, so that computation delays increase by a factor $1/\alpha_n$. Finally, energy and latency feasibility do not guarantee application-level inference accuracy: an admissible deployment must also satisfy $A_s\geq A_{\min}$. Assessing this constraint requires an application-specific AI model and dataset; in particular, different numerical precisions are not assumed to provide identical accuracy.}

\subsection{\rev{Practical Deployment and Scalability of GEO--LEO ISLs}}

\rev{The feasibility of S3 additionally depends on the availability of compatible optical terminals and their platform-level SWaP requirements. Operational LEO--GEO optical relay has been demonstrated by the European Data Relay System (EDRS)~\cite{ESA_EDRS}. However, such links require dedicated optical terminals and PAT mechanisms on both communicating spacecraft. For reference, the TESAT LCT 135 GEO terminal supports $1.8$~Gbit/s and has a mass of approximately $53$~kg, with electrical power consumption of $120$~W during communication and up to $150$~W during acquisition \cite{TESAT_LCT135}. These figures illustrate that the $2$-W ISL transmit/receive power coefficients adopted in Table~\ref{tab:SimParams} are simplified link-energy parameters and do not represent the complete electrical consumption of an operational optical terminal. Consequently, the baseline lifecycle-energy results do not include the full terminal acquisition, tracking, and platform-integration overheads.}

\rev{Link acquisition also affects model-refresh latency. For example, ESA reports an acquisition time of approximately $55$~s for EDRS GEO--LEO optical links \cite{ESA_EDRS_Acquisition}. If an illustrative S3 configuration uses this acquisition delay and an ISL rate of $1.8$~Gbit/s, while retaining the remaining baseline parameters and assuming a single acquisition per refresh cycle, its model-refresh latency increases from approximately $5.02$ to $6.01$~min. The corresponding maximum supported update frequency decreases from $11.95$ to approximately $9.98$~updates/hour. This example is not intended to represent a validated S3 implementation, but illustrates the sensitivity of the feasibility boundaries to terminal-level overheads.}

\rev{At constellation scale, the number of simultaneously available GEO optical terminals, their fields of regard, contact scheduling, and the time required for link acquisition limit the number of LEO satellites that can be served. Multiple LEO refresh requests may therefore introduce queueing delays and competition for GEO communication and computing resources. The present evaluation considers a representative GEO--LEO path with an available ISL and dedicated processing capacity; it does not establish the scalability of a specific multi-satellite deployment. Extending the framework with terminal-concurrency constraints, contact scheduling, and GEO resource sharing is an important direction for future investigation.}

\section{Conclusions}
\label{Sec:Con}
%This paper analyzed the feasibility of distributing the O-RAN control hierarchy across Ground, LEO, and GEO segments to overcome the SWaP-vs-Bandwidth constraints of NTN. Our multi-dimensional sensitivity analysis demonstrates that no single architecture is universally optimal. We proved that the proposed Split-RIC architecture (S2) reduces lifecycle energy by over $90\%$ for data-intensive xApps (inputs $> \qty{85}{\kilo\byte}$), provided the model complexity remains below the hardware ceiling of $\approx \qty{250}{\giga\text{FLOPS}}$. However, for time-critical applications where update urgency exceeds the orbital revisit rate, the Multi-Layer GEO-LEO architecture (S3) is the mandatory solution to bridge connectivity gaps. We conclude that 6G NTN requires a \textit{Flexible Split-RIC} strategy that dynamically migrates functions between the Ground, Edge, and GEO layers based on instantaneous physical constraints. Future work will extend this feasibility framework into a real-time orchestration protocol, utilizing optimization techniques to autonomously determine function placement under time-varying traffic and energy harvesting conditions.

\rev{This paper developed a lifecycle-aware feasibility framework
for distributing O-RAN AI functions across Ground, LEO, and GEO
segments under NTN communication, computation, and power
constraints. By separating space-segment lifecycle energy,
near-real-time control latency, and model-refresh latency, the
analysis shows that no single placement is universally preferable.
For the baseline parameters, S2 becomes more energy-efficient than
S1 when the input telemetry exceeds approximately $83$~kB for a
$1$~GFLOP/inference workload, while at
$\delta^{\mathrm{in}}=5$~MB the corresponding energy crossover
occurs at approximately $60$~GFLOP/inference. These crossovers are
energy-preference boundaries rather than hardware limits, with
computational feasibility separately constrained by the
power-feasible sustained throughput of the spaceborne nodes.}

\rev{The model-refresh analysis further shows that the role of the
GEO-assisted S3 architecture is primarily to reduce dependence on
ground-station availability rather than to minimize energy. Under
the considered baseline, S3 provides a refresh latency of
approximately $5.02$~min and becomes faster than the S2
distinct-pass realization when the ground-contact waiting time
exceeds approximately $2.32$~min; under the favorable same-pass
assumption, the crossover moves to approximately $4.64$~min.
Accordingly, architecture selection should jointly account for
telemetry volume, inference workload, power-feasible computing
capacity, ground-contact intermittency, and the required model
adaptation rate. The derived boundaries are parameter-dependent
decision primitives that can support future SMO-level orchestration
rather than fixed prescriptions for specific xApps. Future work
will extend the framework toward dynamic placement under stochastic
contact and ISL availability, experimentally characterized AI
workloads and hardware performance, and joint consideration of
energy, latency, and inference quality.}

\bibliographystyle{IEEEtran}
\bibliography{IEEEabrv,references}

\begin{IEEEbiography}
[{\includegraphics[width=1in,height=1.25in,clip,keepaspectratio]{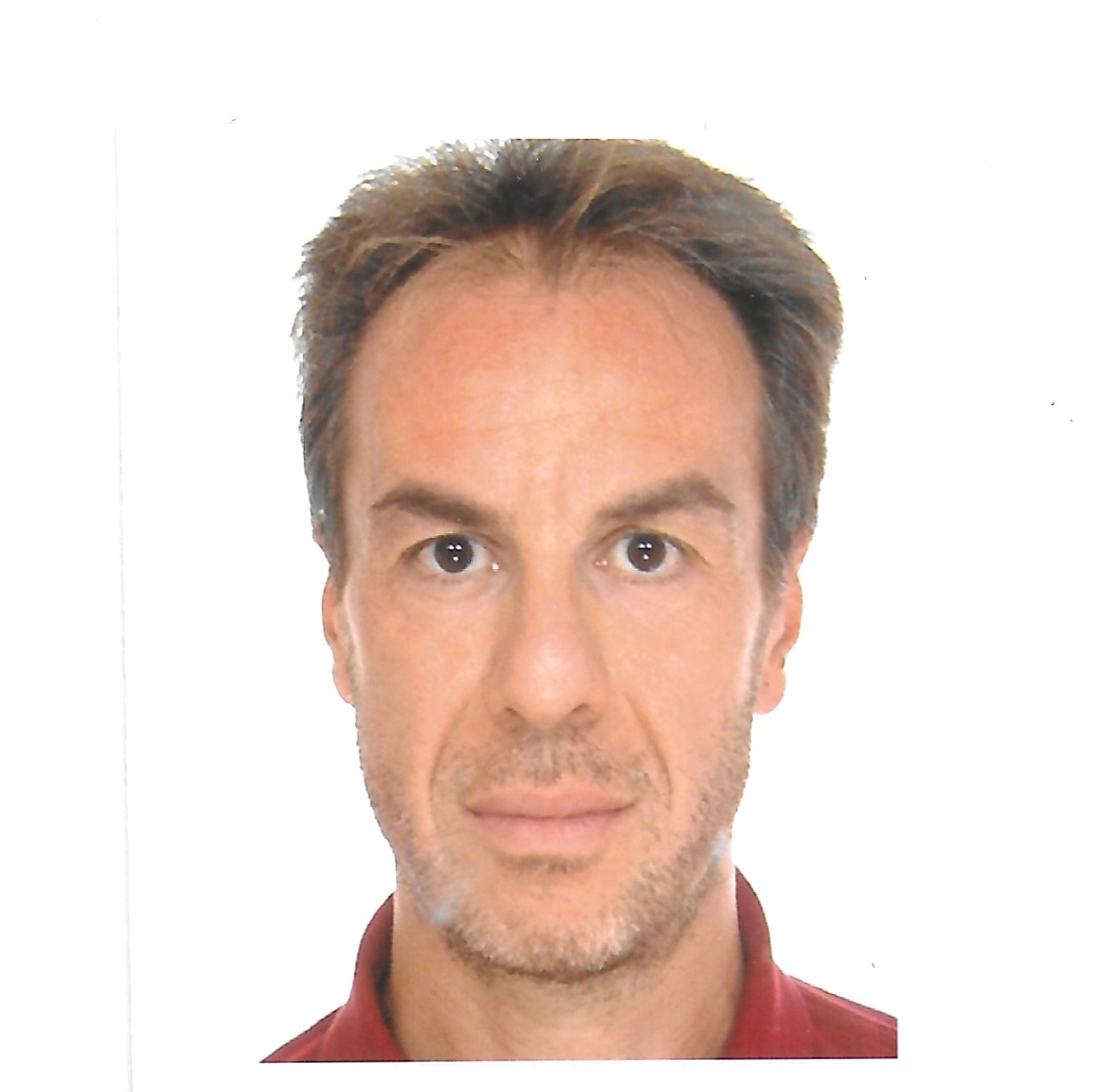}}]
{DANIELE TARCHI}~(Senior Member, IEEE)~was born in Florence, Italy in 1975. He received his M.Sc. degree in Telecommunications Engineering and Ph.D. degree in Informatics and Telecommunications Engineering from the University of Florence, Florence, Italy, in 2000 and 2004, respectively. 

From 2004 to 2010, he was a Research Associate at the University of Florence, Italy. From 2010 until 2019, he was an Assistant Professor and from 2019 to 2024, he was an Associate Professor at the University of Bologna, Italy. Since 2024, he has been an Associate Professor at the University of Florence, Italy. He has authored 170 articles that have been published in international journals and conference proceedings. His research focuses primarily on wireless communications and networks, satellite communications and networks, edge computing, distributed learning, and optimization techniques. He has participated in numerous national and international research projects and collaborates with various research institutes abroad.

Prof. Tarchi has been an IEEE Senior Member since 2012. He is an Editorial Board member for IEEE Transactions on Vehicular Technology, IEEE Open Journal of the Communication Society, and IET Communications. He has been symposium co-chair for IEEE WCNC 2011, IEEE Globecom 2014, IEEE Globecom 2018 and IEEE ICC 2020, and a workshop co-chair at IEEE ICC 2015.
\end{IEEEbiography}

\end{document}